\documentclass[useAMS,usenatbib]{mn2e}
\usepackage{epsfig}
\usepackage{times}
\title[A Newly Discovered Post-AGB Star in M79]{High
Resolution Optical Spectroscopy of a Newly Discovered Post-AGB Star with a
Surprising Metallicity in the Globular Cluster M79}

\author[T. \c{S}ahin and D. L. Lambert]
{T. \c{S}ahin\thanks{E-mail:sahin@astro.as.utexas.edu;dll@astro.as.utexas.edu} and David L. Lambert\\
Department of Astronomy and The W.J.  McDonald Observatory, University of Texas, Austin, TX 78712, USA}

\begin{document}

\date{Accepted .
   Received ;}

\pagerange{\pageref{firstpage}--\pageref{lastpage}} \pubyear{2009}

\maketitle

\label{firstpage}

\begin{abstract}

An abundance analysis based on a high-resolution spectrum
is presented for a newly discovered post-AGB star in the globular
cluster M79. The surprising result is that the iron abundance of the
star is apparently about 0.6 dex less than that of the cluster's red giants as reported
by published studies including a recent high-resolution spectroscopic analysis
by Carretta and colleagues. Abundances relative to iron appear to be the
same for the post-AGB star and the red giants for the 15 common
elements. It is suggested that the explanation for the lower abundances
of the post-AGB star may be that its atmospheric structure differs
from that of a classical atmosphere; the temperature gradient may be
flatter than predicted by a classical atmosphere.

\end{abstract}


\section{Introduction}

\begin{figure}
 \centering
 \includegraphics[width=0.99\columnwidth,angle=0]{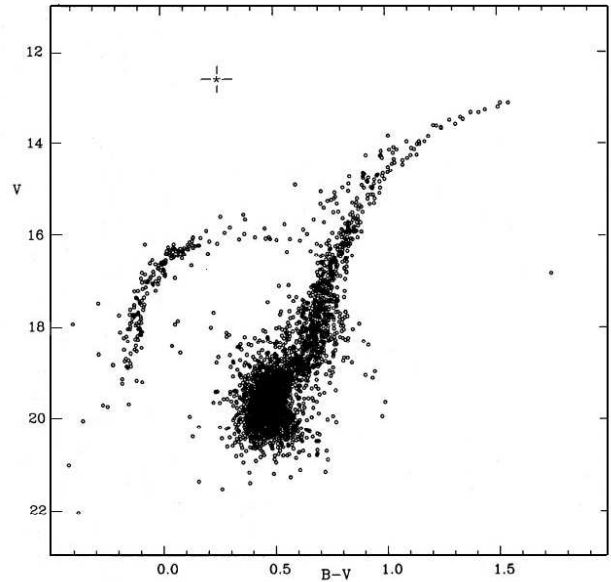}

   \caption{The colour-magnitude diagram of M79 with  photometry  from \citet{ferraro92}.
            The PAGB star is marked by the cross.}
      \label{f_m79_color}
 \end{figure}

\noindent Post-Asymptotic Giant Branch stars (here, PAGB stars) are
low mass stars  evolving between the asymptotic
giant branch (AGB) and the white dwarf cooling track. PAGB stars
evolve from stars with initial masses in the range 0.8$M_\odot$
to 8$M_\odot$. Thanks to mass loss  on the red giant branch
and principally on the  AGB, the PAGB stars are widely considered
to have masses of about $0.8M_\odot$ or less. Evolution from the
cool AGB star to a hot star at the  beginning of the white dwarf cooling track is rapid with
times of 10,000 years thought to be representative.
Gas lost previously by the star is ionized by the hot PAGB central
star to form a planetary nebula.
The dust component of the mass loss may be detected as an infrared excess.
 PAGB stars have
been reviewed by Kwok (1993) and Van Winckel (2003).
\vskip 0.2 cm
\noindent One reason for great interest in PAGB stars is that they have the potential
to provide observational constraints, particularly  through
studies of their chemical compositions, on the complex mix of
evolutionary processes -- nucleosynthesis, mixing and mass loss  --
occurring on the AGB. Interpretation of these constraints for field PAGB stars is
compromised in part  because the
composition and mass of the main sequence, red giant, and AGB progenitor are not
directly known. Such compromises are essentially eliminated by
finding a PAGB star as a member of an open  or globular cluster.
\vskip 0.2 cm
\noindent In this paper, we report on an abundance analysis of the
A-type PAGB star  discovered by
Siegel \& Bond (2009, in preparation) in the globular cluster M79. The location of the star in
the colour-magnitude diagram is shown in Figure~\ref{f_m79_color}. The initial mass of this
star must have been slightly in excess of the mass of stars now at the
main sequence turn-off, say, $M \simeq 0.8M_\odot$.
The star's composition may be referenced to that of the cluster's
red giants, stars for which abundance analyses have been reported.
Comparison of abundances for the PAGB and RGB stars may reveal
changes imposed by the evolution beyond the RGB; such changes are not necessarily
attributable exclusively  to internal nucleosynthesis and dredge-up.
It was in the spirit of
comparing the compositions of the PAGB and RGB stars that we undertook
our analysis. For the RGB stars, we use results kindly provided in advance of publication by
Carretta (2008, private communication).
\vskip 0.2 cm
\noindent PAGB stars because they are  rapidly evolving  are
understandably rare in globular clusters. At spectral types of
F and G, a few luminous variables are known. These are sometimes
referred to as Type II Cepheids. Abundance analyses of  cluster
variables have been reported, for example, for one or two stars in
the clusters M2, M5, M10, and M28 (Gonzalez \& Lambert 1997; Carney,
Fry \& Gonzalez 1998). These stars have ($B-V)_0$ of 0.5-0.6 rather
than the $(B-V)_0=0.28$ of the M79 discovery.
 At even earlier spectral types than A, the
PAGB stars in globular stars are widely referred to as `UV-bright' stars
(see review by Moehler 2001). Three such B-type stars have been
subject to an abundance analysis - see Thompson et al. (2007).
\vskip 0.2 cm
\noindent  In this paper, we present the abundance analysis of the M79 PAGB star and
compare  its composition to that of the
cluster's red giants. Many determinations of the metallicity [Fe/H] of
cluster red giants have given estimates near [Fe/H]$=-1.6$.\footnote{Standard
notation is used for quantities [X] where [X]=$\log$(X)$_{\rm star}-\log$(X)$_\odot$.}
 For example,
Zinn \& West (1984) give [Fe/H]$=-1.69$ and Kraft \& Ivans (2003) give
[Fe/H]$=-1.64$. Recently from high-resolution UVES/FLAMES\footnote{UVES: Ultraviolet and
Visual Echelle Spectrograph, FLAMES: Fibre Large Array Multi Element Spectrograph.} spectra
Carretta and colleagues (2008, private communication) performed an
abundance analysis for 20 elements obtaining
[Fe/H]$=-1.58$ for a sample of ten RGB stars.

\section[]{Observations and Data Reduction}

\begin{figure*}
 \centering
 \includegraphics[width=17cm,angle=360]{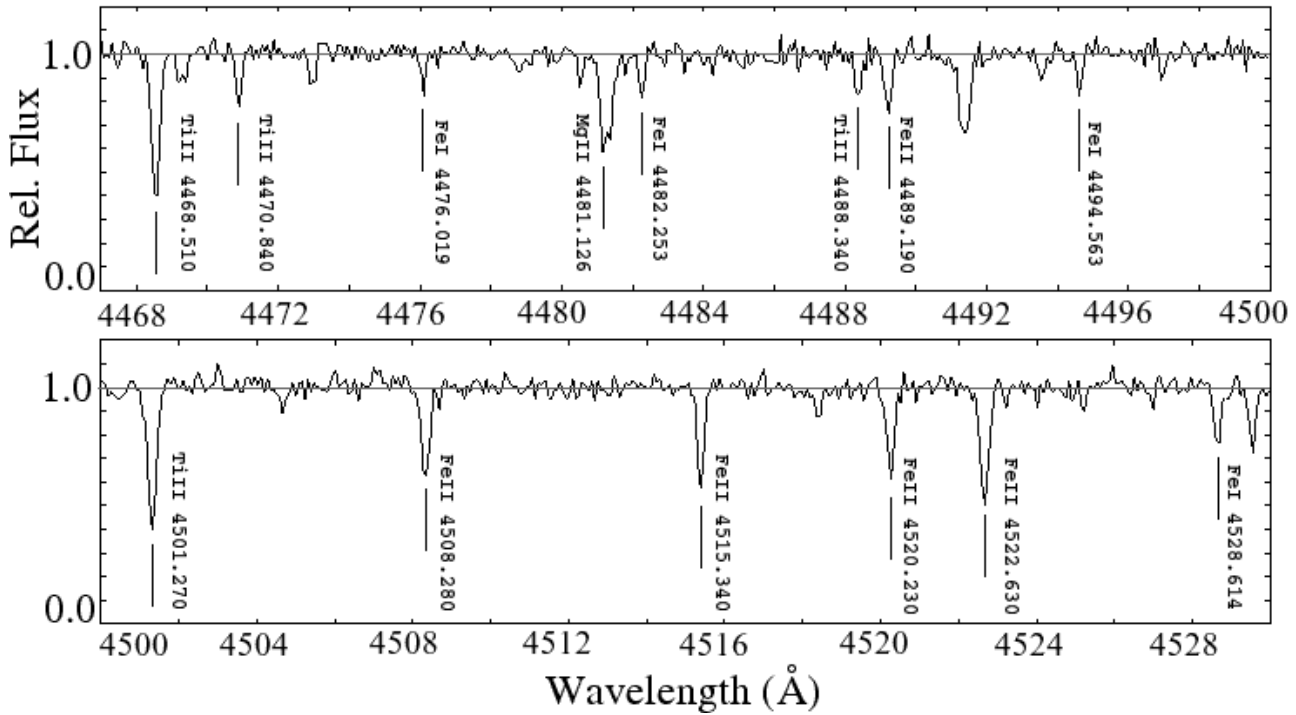}

   \caption{The spectrum for the PAGB star
   over the wavelength regions between 4467-4500 \AA\ (upper panel) and 4500-4530 \AA\
   (lower panel). Selected lines are identified.}
      \label{f_spectrum_part}
 \end{figure*}

\noindent Spectra for the abundance analysis were obtained on five nights between 2008 January
15 and March 3 with the 2.7 meter Harlan J. Smith reflector and its $2dcoud$\'{e}
cross-dispersed \'{e}chelle spectrograph (Tull et al. 1995). The chosen spectral resolving
power was $\lambda/d\lambda \simeq 35,000$ with 3 pixels per resolution element. Full spectral
coverage is provided from 3800 \AA\ to 5700 \AA\  with incomplete but substantial coverage
beyond 5700 \AA; the effective  short and long wavelength limits are set by the useful  S/N
ratio. A ThAr hollow cathode lamp provided the wavelength calibration. Flat-field and bias
exposures completed the calibration files.
\vskip 0.2 cm
\noindent Observations were reduced using the {\sc STARLINK} reduction package {\sc ECHOMOP} (Mills \&
Webb 1994). A series of
30 minute stellar exposures was combined to obtain the final spectrum. The equivalent width (EW) of a line was
measured with the package {\sc DIPSO} using a fitted Gaussian profile for lines weaker than 90 m\AA\ and direct
integration for stronger lines. A section of the final spectrum is shown in Figure~\ref{f_spectrum_part}, among the identified
lines, the strongest line (excluding the Mg\,{\sc ii} blend at 4481 \AA) is the Ti\,{\sc ii} line at 4501.27
\AA\ with an EW of 167 m\AA. The weaker lines of Fe\,{\sc
i} at 4476.019 \AA\ and Ti\,{\sc ii} at 4470.840 \AA\ have EWs of 21 m\AA\ and 38 m\AA, respectively.
\vskip 0.2 cm
\noindent The heliocentric radial velocity measured from the final spectrum is
$+211\pm$5 km s$^{-1}$ with no evidence of a variation greater than
about $\pm$7 km s$^{-1}$ over the observing runs. This velocity
is consistent with the cluster's velocity of $+207.5$ km s$^{-1}$
given by Harris (1996). This agreement between the PAGB star's
velocity and that of the cluster confirms a result
given by Siegel and Bond (2009, in preparation).

\section[]{Abundance Analysis -- The Model Atmospheres}

The abundance analysis was undertaken with model atmospheres
 and the line analysis programme {\sc MOOG} (Sneden 2002).
The models
 drawn from the
{\sc ATLAS9} grid (Kurucz 1993)
 are line-blanketed plane-parallel atmospheres in
Local Thermodynamical Equilibrium (LTE) and hydrostatic equilibrium with flux (radiative plus convective) conservation. MOOG
adopts LTE for the mode of line formation. A model is defined by the parameter set:
effective temperature $T_{\rm eff}$, surface gravity $g$, chemical composition as
represented by metallicity [Fe/H] and  all models  are computed for a
microturbulence $\xi$ = 2 km s$^{-1}$. A model defined by the parameter set is
fed to MOOG except that $\xi$ is determined from the spectrum and  not set to
the canonical 2 km s$^{-1}$ assumed for the model atmosphere.
\vskip 0.2 cm
\noindent Several methods are available for obtaining estimates of
the atmospheric parameters from photometry and
spectroscopy. Most methods are sensitive to both $T_{\rm eff}$ and
$\log g$ and, therefore, provide loci in the $T_{\rm eff}$ versus
$\log g$ plane.
We discuss several methods in an attempt to  find  consistent values for the
effective temperature and gravity.

\subsection{Photometry}

\noindent Bond (2005, see also Siegel \& Bond 2005) developed a photometric system with
one of several aims being the detection of `stars of high luminosity
in both young (yellow supergiants) and old (post-AGB stars)
populations'. The system combines the Thuan-Gunn $u$ filter
with the Johnson-Kron-Cousins $B$, $V$, and $I$ filters. Stellar
parameters $T_{\rm eff}$ and gravity $g$ are obtainable from
the colour-colour diagrams ($u-B$) versus ($B-V$) and
$(u-B)-(B-V)$ versus $(V-I)$. Bond provides calibration diagrams
for metallicities [Fe/H] = 0 and $-2$.
\vskip 0.2 cm
\noindent Siegel (2008, private communication) reports the magnitudes of
the M79 PAGB star to be $u$=13.803, $B$=12.480, $V$=12.203 and
$I$=11.744. M79 is reddened by a negligible amount: $E(B-V)= 0.01$ magnitudes
(Heasley et al. 1986; Ferraro et al. 1992, 1999), a correction that  is ignored here.
This photometry and the [Fe/H] $=-2$ grid provides the following
estimates
\vskip 0.2 cm
\noindent ($u-B$) vs ($B-V$): $T_{\rm eff}=6400$K and $\log g=1.2$ cgs units. \\
\noindent $(u-B)-(B-V)$ vs $(V-I)$: $T_{\rm eff}=6300$K and $\log g=0.8$ cgs units.\\
\vskip 0.2 cm
\noindent The mean results are $T_{\rm eff}$=6350K and $\log g=1.0$ cgs units.
Our abundance analysis indeed suggests that [Fe/H] $\simeq -2$.
\vskip 0.2 cm
\noindent Analyses of the cluster's red giants have, however, found [Fe/H] $\simeq -1.6$. Interpolation,
necessarily crude given grids at only [Fe/H] = 0 and $-2$, suggests that for
[Fe/H] $=-1.6$, the $T_{\rm eff}$  and $\log g$ are increased by 450 K
and 0.8 dex, respectively.
\vskip 0.2 cm
\noindent A calibration of the $(B-V)$ colour by Sekiguchi \& Fukugita (2000) gives
$T_{\rm eff}=6625$K (their equation 2) but the parameter space near [Fe/H]$=-2$ and
$\log g=1.5$ is poorly represented by calibrating stars.

 \begin{figure}
 \centering
 \includegraphics[width=0.99\columnwidth,angle=0]{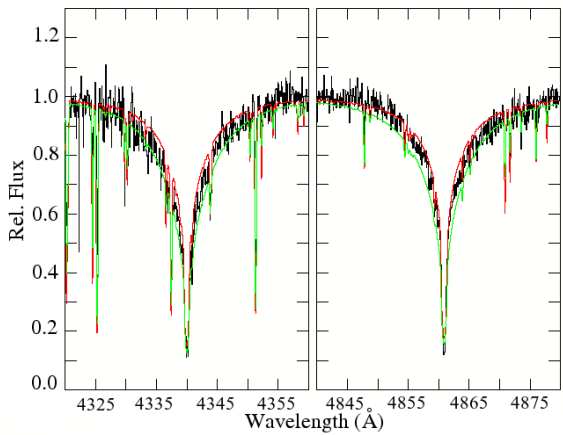}

   \caption{Observed and model line profiles for H$\beta$ and H$\gamma$. The theoretical profiles have been generated
for a surface gravity $\log g = 1.0$ dex and a metallicity [Fe/H]$=-2$. The red and green lines show the theoretical
profiles for $T_{\rm eff}$= 6000 and 6500 K, respectively.}
      \label{f_balmer_profiles}
 \end{figure}

\subsection{Balmer lines}

\noindent The Balmer lines at temperatures around 6300 K are
sensitive to $T_{\rm eff}$ with a  weak dependence on
gravity.
The synthetic spectra for the Balmer lines H${\rm \beta}$ and H${\rm \gamma}$ have been computed
with SYNTHE which is a suite of different programs , called one after the other by an input script
with the purpose of producing a synthetic spectrum. These profiles were convolved with a
gaussian profile in DIPSO to simulate the instrumental broadening. At $T_{\rm eff} <
6000$ K, the theoretical  H$\beta$ and H$\gamma$ profiles are narrower than the observed
profiles for all plausible values of the surface gravity. At $T_{\rm eff} >6500$ K, the
theoretical profiles are broader than the observed profiles.  Acceptable fits are found
for $T_{\rm eff}= 6250$ K. A change of greater than  $\pm200$K in $T_{\rm
eff}$ provides an unsatisfactory fit of theoretical to observed profiles. A gravity range
of $\log g$ = 1.0 to 2.0 (and probably greater) does not impair the fit. In
Figure~\ref{f_balmer_profiles}, we show the observed  profiles with theoretical line
profiles for $T_{\rm eff}$=6000 K and 6500 K for $\log g =1.0$ and [Fe/H] $= -2$.

%%%%%%%%%%%%%%%%%%%%%%%%%%%%%%%%%%%%%%%%%%%%%%%%%%%%%%%%%%%%%%%%%%%%%
%%%%%%%%%%% FIGURE PRESENTING TEFF and VT DETERMINATION %%%%%%%%%%%%%
%%%%%%%%%%%%%%%%%%%%%%%%%%%%%%%%%%%%%%%%%%%%%%%%%%%%%%%%%%%%%%%%%%%%%

 \begin{figure}
 \includegraphics[width=0.99\columnwidth,height=85mm,angle=0]{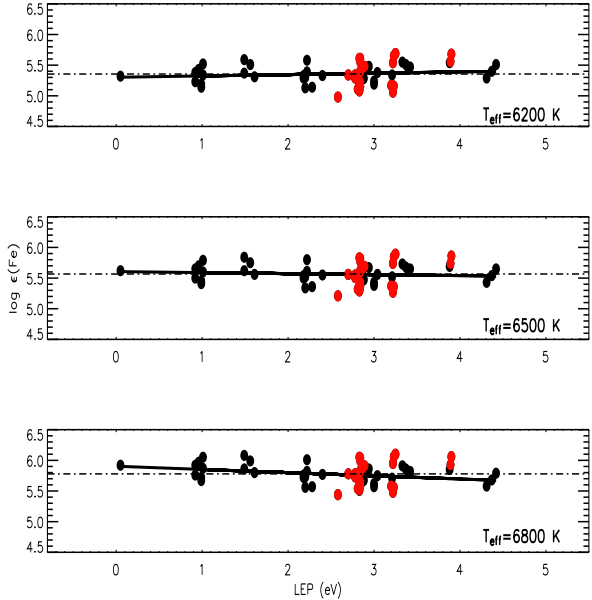}

%   \vspace{3.5cm}
\caption{Iron abundances from individual Fe\,{\sc i} (black filled circles)
 and Fe\,{\sc ii} (red filled circles) lines versus a line's
lower excitation potential (LEP). Results are shown for three effective temperatures.
The solid line in each plot is the least-squares fit to the abundances from the Fe\,{\sc i}
lines. The non-zero slope of this line for $T_{\rm eff}=6200$K and 6800K suggests that these
values are too cool and too hot, respectively. The Fe\,{\sc ii} lines in the mean are forced to provide the
same Fe abundance as the Fe\,{\sc i} lines by means of a different choice of surface gravity for
each effective temperature (see text). The microturbulence velocity is 3.4 km s$^{-1}$.}
      \label{f_excitation}
 \end{figure}

%%%%%%%%%%%%%%%%%%%%%%%%%%%%%%%%%%%%%%%%%%%%%%%%%%%%%%%%%%%%%%%%%%%%%
%%%%%%%%%%%%%%%%%%%%%%%%%%%%%%%%%%%%%%%%%%%%%%%%%%%%%%%%%%%%%%%%%%%%%
%%%%%%%%%%%%%%%%%%%%%%%%%%%%%%%%%%%%%%%%%%%%%%%%%%%%%%%%%%%%%%%%%%%%%

\subsection{Spectroscopy - Fe\,{\sc i} and Fe\,{\sc ii} lines}

\noindent A selection of 42 Fe\,{\sc i} lines was measured
with lower excitation potentials (LEP) ranging from
0 to 4.4 eV and EWs of up to 170 m\AA\ but only three lines
have EW greater than 100 m\AA.
Measured Fe\,{\sc i} and 25  Fe\,{\sc ii} lines are
listed in Table 1. The $gf$-values are taken from the
recent critical compilation by Fuhr \& Wiese (2006).
\vskip 0.2 cm
\noindent In the limit that a line selection contains only weak lines, the
$T_{\rm eff}$ is found by imposing the condition that the
derived abundance be independent of the LEP. In the limit that all
lines have the same LEP and a similar wavelength, the microturbulence $\xi$
is found by requiring that the derived abundance be independent of
the EW. For our sample of Fe\,{\sc i} lines, these two
conditions must be imposed simultaneously.
\vskip 0.2 cm
\noindent No species other than Fe\,{\sc i}
provides a sample of lines spanning an adequate range in LEP and
EW to provide additional estimates of both $T_{\rm eff}$ and $\xi$.
Solutions for $T_{\rm eff}$ and $\xi$ are not very sensitive
to the adopted surface gravity.
To illustrate the sensitivity to $T_{\rm eff}$, we  show in
Figure~\ref{f_excitation} the abundance-LEP relations for 6200 K, 6500 K,  and 6800 K,
i.e., $\pm300$ K around the best value. It is to be noted that the
slopes of the least-squares fitted  relations are for 6200 K and 6800 K slight but possibly acceptable as
different from zero.
\vskip 0.2 cm

 \begin{figure}
 \centering

 \includegraphics[width=0.99\columnwidth,height=85mm,angle=90]{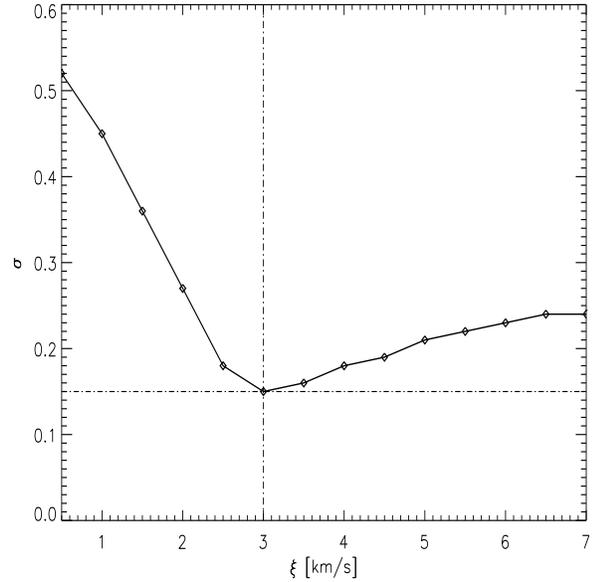}
% \includegraphics[width=0.99\columnwidth,height=85mm,angle=0]{PAPER_microturbulence_feI.ps}
%   \caption{Microturbulence velocity from individual Fe\,{\sc i} (black filled circles) lines versus a line's reduced equivalent width
%($log (W_{\lambda}/\lambda)$). The red lines in each plot are the least-squares fits to the abundances from FeI lines for 
%$T_{\rm eff}=6000$ K (bottom red line) and 6600 K (top red line) models. The gravity is same as consensus model, $\log g=1.0$. The dotted-black
%lines show the mean abundances for $T_{\rm eff}=6600$, 6300, and 6000 K from top to bottom. The thick black lines show the least-squares fits for the consensus model.}
\caption{The standard deviation of the Fe abundance from the suite of Fe\,{\sc i}
lines as a function of the microturbulence $\xi$ for a model
with $T_{\rm eff}=6300$ K, $\log$ g = 0.8, and [Fe/H] $=-2$.}
      \label{f_micro}
 \end{figure}

\noindent The microturbulence is determined separately from the Fe\,{\sc i} and
Ti\,{\sc ii} lines. For a given model, we compute the dispersion
in the Fe (or Ti) abundances over a range in the $\xi$. Figure~\ref{f_micro}
shows the dispersion $\sigma$ for Fe for $\xi$ from 0.5 to
7 km s$^{-1}$. A minimum value of $\sigma$ is reached at
$\xi$ = 3.0 km s$^{-1}$ and a range of 2.2 km s$^{-1}$ to 5.0 km s$^{-1}$
covers the $\pm$50$\%$ range about the minimum value of $\sigma$.
 A value slightly less than $\xi = 5$ km s$^{-1}$ is a firm upper
limit because at higher values the predicted line widths exceed the
observed widths, even if the macroturbulence is put at the unlikely
value of 0 km s$^{-1}$. A similar exercise with the Ti\,{\sc ii} lines
gives $\xi = 3.8$ km s$^{-1}$. We adopt a mean value of 3.4 km s$^{-1}$.
These results are for a model with $T_{\rm eff}$ = 6300 K, $\log$ g = 0.8,
and [Fe/H] =$-2$. Tests show that the derived $\xi$ is insensitive to
these parameters over quite a wide range. Our abundances are primarily
based on weak lines and are, thus, insensitive to the precise value
of $\xi$.
\vskip 0.2 cm
\noindent A third condition provides an
estimate of the gravity. This is the familiar requirement that
Fe\,{\sc i} and Fe\,{\sc ii} lines provide a single value of the
Fe abundance. Obviously, this condition of ionization
equilibrium provides a locus in the
temperature-gravity plane running from low $T_{\rm eff}$ and low $g$ to high
$T_{\rm eff}$ and high $g$ with the iron abundance increasing along this
locus.

\subsection{Ionization equilibria for Mg and Cr}

Often, iron is the primary and occasionally the sole element used via ionization
equilibrium to provide a $T_{\rm eff} - \log g$ locus
which with an independent estimate of $T_{\rm eff}$ is used to
obtain an estimate of $\log g$. The primary reason for iron's
supremacy  is that it provides a plentiful
collection of both neutral and ionized
 lines: here, 42 Fe\,{\sc i} and 25 Fe\,{\sc ii} lines.
\vskip 0.2 cm
\noindent Our spectrum provides other elements with lines from the
neutral and singly-ionized atoms, although much less well
represented than iron. The elements in question are Mg and Cr
for which the available lines are listed in Table 3. (The sources of the
$gf$-values for these lines are identified in the subsequent discussion of elemental
abundances.) Figure~\ref{f_logg_teff} shows that Mg, Fe and Cr loci in the $T_{\rm eff} -
\log g$ plane.

\subsection{Mass and Luminosity}

An aspect of globular cluster membership is that one can
estimate a star's surface gravity from the
known luminosity, effective temperature and a constraint on the
estimated  stellar mass. Combining the relations
$L \propto R^2T_{\rm eff}^4$ and $g \propto M/R^2$, one
obtains

\begin{equation}
\hskip 1.5 cm log L/L_\odot = log M/M_\odot  + 4log T_{\rm eff} - log g - 10.61
\end{equation}

\vskip 0.2 cm
\noindent The absolute visual magnitude is $M_V = -3.37$.
This with bolometric corrections from Fiorella Castelli (2008, private
communication)  provides estimates\footnote{http://wwwuser.oat.ts.astro.it/castelli/colors/bcp.html}
of $L/L_\odot$ as weak functions of $T_{\rm eff}$, $\log g$, and
[Fe/H]: for example, $T_{\rm eff}$ = 6500K, $\log g$=1.5 and
[Fe/H]= -2.0 give $L/L_\odot$ = 1740.
This absolute luminosity is consistent with the range
$L/L_\odot$ = 1070 to 1862 reported by Gonzalez \& Lambert (1997) from five RV Tauri
variables in other globular clusters and the theoretical characteristic that PAGB stars evolve at approximately constant
luminosity.
 The mass of the
PAGB star is less than 0.8$M_\odot$ and greater than the mass of
white dwarfs in the cluster, say 0.5$M_\odot$ from mass estimates
of the PAGB and white dwarfs in globular clusters (0.53$M_\odot$ is the typical PAGB
remnant mass in NGC 5986 (Alves, Bond, \& Onken 2001) and  0.50$\pm$0.02$M_\odot$ is the average
mass of white dwarfs in nearby GCs and the halo field (Alves, Bond, \& Livio 2000 and
references therein). Adopting
a mass of 0.6$M_\odot$, $T_{\rm eff}$=6500 K, and $L/L_\odot$ = 1740, we obtain the
 $\log g$=1.18. Extending this procedure to other
$T_{\rm eff}$   gives a
locus in the $T_{\rm eff},\log g$ plane but one which has a quite different
slope to those from other indicators (Figure~\ref{f_logg_teff}).

\subsection{The Atmospheric Parameters}

\begin{figure}
 \centering
 \includegraphics[width=1.0\columnwidth,height=65mm]{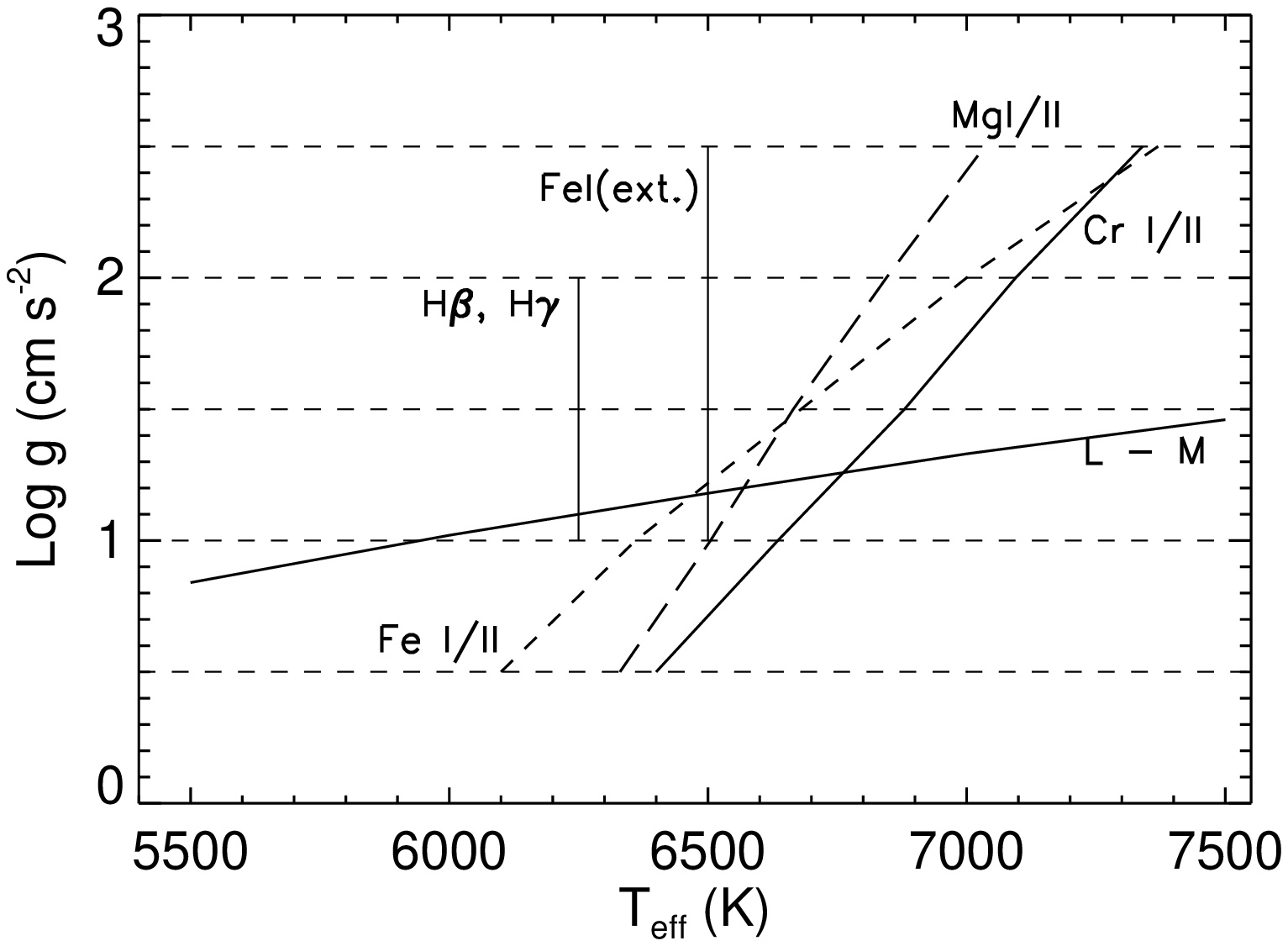}

   \caption{The $T_{\rm eff}$ versus $\log g$ plane showing the various loci discussed
in the text. The loci include those derived from the fit to the Balmer line profiles H$\beta$ and
H$\gamma$, the mass and luminosity estimates (L-M),
the excitation of the Fe\,{\sc i} lines, (FeI(ext))
 and the imposition of
ionization equilibrium  for Mg (dashed line),  Cr (solid line), and
Fe (dashed line). }
      \label{f_logg_teff}
 \end{figure}

\noindent Figure~\ref{f_logg_teff} shows the loci discussed above. Convergence  of the
loci suggests adoption of the parameters ($T_{\rm eff}$ in K, $\log g$ in cgs) =
(6300,0.8). The microturbulence $\xi = 3.4$ km s$^{-1}$ is adopted from the Fe\,{\sc i}
and Ti\,{\sc ii} lines analysis. The corresponding iron abundance is $\log\epsilon$(Fe) =
5.42 or [Fe/H] = $-2.03$  for the solar Fe abundance of $\log\epsilon$(Fe)=7.45 (Asplund
et al. 2005). We refer to this model as the consensus choice.
\vskip 0.2 cm
\noindent With these parameters, Fe ionization equilibrium is satisfied by design. Mg, and
Cr ionization equilibria are quite well met: the abundance differences in the sense of
neutral minus ionized lines is -0.1 dex for Mg and -0.2 dex for Cr. Also, there is a
satisfactory fit to the Balmer line profiles, the excitation of the Fe\,{\sc i} lines, and
the $uBVI$ photometry.
\vskip 0.2 cm
\noindent One is struck immediately by the fact that the consensus model
yields a [Fe/H] that is lower by about 0.6 dex than the abundance previously derived for this
cluster from its red giants (see Introduction).
No modern abundance determination known to us
has obtained a result near [Fe/H] $\simeq -2.0$.  Adjustments for different assumptions
about the solar Fe abundance will not reconcile previous and our results.
\vskip 0.2 cm
\noindent As  part of a preliminary exploration of ways to reconcile the
composition of the PAGB star with that of the red giants, we
report in Table 3 abundance analyses of the PAGB star for four
model atmospheres falling along the ionization equilibrium track for
iron: ($T_{\rm eff},\log g)$ = (6300,0.80), (6500,1.18), (6800,1.67),
and (7000,1.98) where the final model returns essentially the Fe abundance reported
for the cluster from its red giants. These abundances are obtained using the line selections described
in the next section.
\vskip 0.2 cm
\noindent In Table 3, the quantities $\log\epsilon$(X) and [X/Fe]
reported by Carretta are given in
columns two and three.
In the next eight columns,
we give our abundances and [X/Fe] computed from the solar abundances
given by Asplund et al. (2005) which are given in the final column.

%%%%%%%%%%%%%%%%%%%%%%%%%%%%%%%%%%%%%%%%%%%%%%%%%%%%%%%%%%%%%%%%%%%%%
%%%%%%%%%%%%%%%%%%%%%%%%%%%%%%%%%%%%%%%%%%%%%%%%%%%%%%%%%%%%%%%%%%%%%

%%%%%%%%%%%%%%%%%%%%%%%%%%%%%%%%%%%%%%%%%%%%%%%%%%%
%%%%%%%%%%%%%  Fe LINES USED IN THE ANALYSIS %%%%%%
%%%%%%%%%%%%%%%%%%%%%%%%%%%%%%%%%%%%%%%%%%%%%%%%%%%

  \begin{table*}
     \caption[]{Fe I and Fe II lines used in the analysis and corresponding abundances for the consensus model.}
        \label{}
    $$
        \begin{array}{@{}cc@{}r@{}c@{}rr||cc@{}r@{}c@{}rr@{}}
           \hline
           \hline
 $Species$&\lambda &  $EW$\,\, &\log\epsilon(Fe)\,\,& $LEP$   &\log(gf)   & $Species$&\lambda &  $EW$ \,\,&\log\epsilon(Fe)\,\,& $LEP$   &\log(gf)\\
        & ($\AA$)& (m$\AA$)& (dex)          & ($eV$)  &           & 	& ($\AA$)& (m$\AA$)& ($dex$)	       & ($eV$)  &	\\
           \hline

%4294.140^{\dagger}& Fe$\,{\sc i}$&1.485& -1.11&  197& 7.15
%Fe$\,{\sc i}$ &5497.516  & 14 & 5.46&1.011&-2.85   \\ not a real line

%NOT: 4383.54 and 4404.760 show the highest deviations as -0.63 and -0.57 respectively.

Fe$\,{\sc i}$&4383.540    &167 & 5.1&1.49&  0.20 &   Fe$\,{\sc i}$ &5434.524  & 51 & 5.4&1.01&-2.12   \\
Fe$\,{\sc i}$&4404.760    &138 & 5.1&1.56& -0.14 &   Fe$\,{\sc i}$ &5446.917  & 53 & 5.2&0.99&-1.91   \\
Fe$\,{\sc i}$&4427.312    &42  & 5.4&0.05& -3.04 &   Fe$\,{\sc i}$ &5455.630  & 70 & 5.6&1.01&-2.09   \\
Fe$\,{\sc i}$&4442.390    &29  & 5.4&2.20& -1.26 &   Fe$\,{\sc i}$ &5569.620  & 23 & 5.5&3.42&-0.49   \\
Fe$\,{\sc i}$&4447.722    &28  & 5.5&2.22& -1.34 &   Fe$\,{\sc i}$ &5572.840  & 33 & 5.5&3.40&-0.28   \\
Fe$\,{\sc i}$&4459.121    &30  & 5.4&2.18& -1.28 &   Fe$\,{\sc i}$ &5586.755  & 49 & 5.6&3.37&-0.14   \\
Fe$\,{\sc i}$&4466.552    &24  & 5.2&2.83& -0.60 &   Fe$\,{\sc i}$ &5615.652  & 72 & 5.6&3.33& 0.05   \\
Fe$\,{\sc i}$&4476.019    &21  & 5.4&2.85& -0.82 &   Fe$\,{\sc i}$ &6494.980  & 24 & 5.4&2.40&-1.27   \\
Fe$\,{\sc i}$&4482.253    &31  & 5.7&2.22& -1.48 &   Fe$\,{\sc ii}$&4173.490  & 73 & 5.1&2.58&-2.61   \\
Fe$\,{\sc i}$&4494.563    &26  & 5.2&2.20& -1.14 &   Fe$\,{\sc ii}$&4303.200  & 96 & 5.4&2.70&-2.61   \\
Fe$\,{\sc i}$&4528.614    &61  & 5.4&2.18& -0.82 &   Fe$\,{\sc ii}$&4385.380  & 89 & 5.4&2.78&-2.58   \\
Fe$\,{\sc i}$&4871.318    &54  & 5.4&2.87& -0.36 &   Fe$\,{\sc ii}$&4416.817  & 92 & 5.4&2.78&-2.60   \\
Fe$\,{\sc i}$&4872.138    &33  & 5.4&2.88& -0.57 &   Fe$\,{\sc ii}$&4472.920  & 22 & 5.5&2.84&-3.53   \\
Fe$\,{\sc i}$&4878.211    &18  & 5.4&2.89& -0.89 &   Fe$\,{\sc ii}$&4489.190  & 58 & 5.4&2.83&-2.97   \\
Fe$\,{\sc i}$&4957.593    &134 & 5.7&2.81&  0.23 &   Fe$\,{\sc ii}$&4508.280  &117 & 5.5&2.86&-2.35   \\
Fe$\,{\sc i}$&5001.860    &28  & 5.6&3.88& -0.01 &   Fe$\,{\sc ii}$&4515.340  & 92 & 5.2&2.84&-2.36   \\
Fe$\,{\sc i}$&5041.756    &20  & 5.5&1.49& -2.20 &   Fe$\,{\sc ii}$&4520.230  & 87 & 5.4&2.81&-2.62   \\
Fe$\,{\sc i}$&5049.819    &16  & 5.2&2.28& -1.36 &   Fe$\,{\sc ii}$&4522.630  &128 & 5.3&2.84&-1.99   \\
Fe$\,{\sc i}$&5051.634    &18  & 5.5&0.92& -2.80 &   Fe$\,{\sc ii}$&4555.890  & 97 & 5.2&2.83&-2.25   \\
Fe$\,{\sc i}$&5139.463    &48  & 5.5&2.94& -0.51 &   Fe$\,{\sc ii}$&4576.330  & 78 & 5.6&2.84&-2.92   \\
Fe$\,{\sc i}$&5171.596    &62  & 5.7&1.49& -1.79 &   Fe$\,{\sc ii}$&4582.840  & 70 & 5.7&2.84&-3.06   \\
Fe$\,{\sc i}$&5191.455    &32  & 5.4&3.04& -0.55 &   Fe$\,{\sc ii}$&4620.510  & 29 & 5.3&2.83&-3.19   \\
Fe$\,{\sc i}$&5192.344    &34  & 5.3&3.00& -0.42 &   Fe$\,{\sc ii}$&4629.340  &101 & 5.2&2.81&-2.26   \\
Fe$\,{\sc i}$&5194.942    &29  & 5.6&1.56& -2.09 &   Fe$\,{\sc ii}$&4666.750  & 46 & 5.7&2.83&-3.37   \\
Fe$\,{\sc i}$&5266.555    &33  & 5.3&3.00& -0.39 &   Fe$\,{\sc ii}$&4731.440  & 50 & 5.6&2.89&-3.13   \\
Fe$\,{\sc i}$&5270.356    &68  & 5.4&1.61& -1.34 &   Fe$\,{\sc ii}$&5197.560  & 91 & 5.2&3.23&-2.05   \\
Fe$\,{\sc i}$&5324.180    &50  & 5.4&3.21& -0.10 &   Fe$\,{\sc ii}$&5234.620  &112 & 5.6&3.22&-2.21   \\
Fe$\,{\sc i}$&5367.500    &26  & 5.6&4.42&  0.44 &   Fe$\,{\sc ii}$&5264.810  & 34 & 5.7&3.23&-3.23   \\
Fe$\,{\sc i}$&5369.962    &27  & 5.5&4.37&  0.54 &   Fe$\,{\sc ii}$&5276.000  &109 & 5.2&3.20&-1.90   \\
Fe$\,{\sc i}$&5383.369    &30  & 5.3&4.31&  0.65 &   Fe$\,{\sc ii}$&5325.560  & 40 & 5.1&3.22&-2.57   \\
Fe$\,{\sc i}$&5397.128    &78  & 5.3&0.92& -1.99 &   Fe$\,{\sc ii}$&5534.890  & 66 & 5.8&3.25&-2.86   \\
Fe$\,{\sc i}$&5404.200    & 26 & 5.5&4.44& 0.52  &   Fe$\,{\sc ii}$&6238.430  & 24 & 5.6&3.89&-2.75   \\
Fe$\,{\sc i}$&5405.775    & 63 & 5.3&0.99&-1.84  &   Fe$\,{\sc ii}$&6456.390  & 72 & 5.7&3.90&-2.19   \\
Fe$\,{\sc i}$&5429.696    & 86 & 5.5&0.96&-1.88  &                 &          &    &     &     &        \\
\hline
\hline
        \end{array}
    $$
%\begin{list}{}{}
%\item  \hskip 1.0 cm $^{\mathrm{\dagger}}$ The strong lines are not included in the analysis.
%When \item  \hskip 1.3 cm these strong lines are included, iron abundance is found to be
%log$\epsilon$(Fe\,{\sc i})=5.79$\pm$0.23, log$\epsilon$(Fe\,{\sc ii})=5.82$\pm$0.26, and  $\xi_{t}$=3.8 versus
%3.7 km s$^{-1}$. \item  \hskip 1.3 cm The relatively higher standard deviation should be noted.
%\item  \hskip 1.0 cm $^{\mathrm{*}}$ F\"{u}hr \& Wiese (2006)

%\end{list}
  \end{table*}

%%%%%%%%%%%%%%%%%%%%%%%%%%%%%%%%%%%%%%%%%%%%%
%%%%%% LINE LIST FOR ALL ELEMENTS %%%%%%%%%%%
%%%%%%%%%%%%%%%%%%%%%%%%%%%%%%%%%%%%%%%%%%%%%

%  O$\,{\sc i}  $& 9260.806 & SS  &\leq7.8&10.74 & -0.24  &    
%  O$\,{\sc i}  $& 9260.848 & SS  &\leq7.8&10.74 &  0.11  &    
%  O$\,{\sc i}  $& 9260.936 & SS  &\leq7.8&10.74 &  0.00  &    

%  O$\,{\sc i}  $& 9262.582 & SS  &\leq7.8&10.74 & -0.37  &    
%  O$\,{\sc i}  $& 9262.670 & SS  &\leq7.8&10.74 &  0.22  &    
%  O$\,{\sc i}  $& 9262.776 & SS  &\leq7.8&10.74 &  0.43  &    

  \begin{table*}
     \caption[]{Equivalent width measurements from the optical spectra. The corresponding abundances are
     presented for the consensus model.}
        \label{}
    $$
        \begin{array}{l@{}cr@{}rrr||l@{}cr@{}rcr@{}}
           \hline
          \hline
$Species$ & \lambda & $EW$\,\,\, & \log\epsilon(X) & $LEP$ & \log(gf) & $Species$ & \lambda & $EW$\,\,\, & \log\epsilon(X) & $LEP$ & \log(gf)  \\
        &  ($\AA$) & (m$\AA$)& ($dex$)     &($eV$)&       &        &($\AA$)&(m$\AA$)& ($dex$)           &($eV$)&    \\
\hline
C$\,{\sc i}  $& 9061.440 & SS  &\leq5.7&7.48  &	-0.35 &   Ti$\,{\sc ii}$& 4468.510 & 166 &  3.0 &1.13& -0.62 \\
C$\,{\sc i}  $& 9062.490 & SS  &\leq5.7&7.48  &	-0.46 &   Ti$\,{\sc ii}$& 4470.840 & 38  &  3.1 &1.17& -2.28 \\
C$\,{\sc i}  $& 9078.290 & SS  &\leq5.7&7.48  &	-0.58 &   Ti$\,{\sc ii}$& 4488.340 & 41  &  3.4 &3.12& -0.82 \\
O$\,{\sc i}  $& 8446.247 & SS  &  8.0 & 9.52 & -0.46  &   Ti$\,{\sc ii}$& 4501.270 & 167 &  3.1 &1.12& -0.75 \\
O$\,{\sc i}  $& 8446.359 & SS  &  8.0 & 9.52 &  0.24  &   Ti$\,{\sc ii}$& 4529.480 & 61  &  3.5 &1.57& -2.03 \\
O$\,{\sc i}  $& 8446.758 & SS  &  8.0 & 9.52 &  0.01  &   Ti$\,{\sc ii}$& 4533.970 & 212 &  3.9 &1.24& -0.77 \\
O$\,{\sc i}  $& 9260.863 & SS  &\leq8.0&10.74 & -0.04 &   Ti$\,{\sc ii}$& 4563.770 & 161 &  3.3 &1.22& -0.96 \\
O$\,{\sc i}  $& 9262.676 & SS  &\leq8.0&10.74 &  0.09 &   Ti$\,{\sc ii}$& 4571.960 & 185 &  3.6 &1.57& -0.52 \\
Na$\,{\sc i} $& 8183.255 & SS  &\leq4.0&2.10  &  0.24 &   Ti$\,{\sc ii}$& 4708.650 & 35  &  3.0 &1.24& -2.21 \\
Na$\,{\sc i} $& 8194.790 & SS  &\leq4.0&2.10  & -0.46 &   Ti$\,{\sc ii}$& 4779.980 & 70  &  3.3 &2.05& -1.37 \\
Mg$\,{\sc i} $& 4057.480 & 32  &  5.8 &4.35  & -0.90  &   Ti$\,{\sc ii}$& 4798.530 & 30  &  3.0 &1.08& -2.43 \\
Mg$\,{\sc i} $& 4167.230 & 26  &  5.6 &4.35  & -0.75  &   Ti$\,{\sc ii}$& 4805.090 & 78  &  3.2 &2.06& -1.12 \\
Mg$\,{\sc i} $& 4703.000 & 66  &  5.8 &4.35  & -0.44  &   Ti$\,{\sc ii}$& 4874.010 & 24  &  3.1 &3.10& -0.79 \\
Mg$\,{\sc i} $& 5183.620 & 201 &  5.7 &2.72  & -0.17  &   Ti$\,{\sc ii}$& 5129.160 & 50  &  3.0 &1.89& -1.40 \\
Mg$\,{\sc i} $& 5528.410 & 61  &  5.8 &4.35  & -0.50  &   Ti$\,{\sc ii}$& 5154.070 & 66  &  3.4 &1.57& -1.92 \\
Mg$\,{\sc i} $& 8806.760 & 153 &  6.2 &4.35  & -0.13  &   Ti$\,{\sc ii}$& 5185.900 & 48  &  2.9 &1.89& -1.35 \\
Mg$\,{\sc ii}$& 4481.126 & SS  &  5.9 &8.86  &  0.75  &   Ti$\,{\sc ii}$& 5188.690 & 125 &  3.3 &1.58& -1.22 \\
Mg$\,{\sc ii}$& 4481.150 & SS  &  5.9 &8.86  & -0.55  &   Ti$\,{\sc ii}$& 5211.544 & 28  &  3.2 &2.59& -1.36 \\
Mg$\,{\sc ii}$& 4481.325 & SS  &  5.9 &8.86  &  0.59  &   Ti$\,{\sc ii}$& 5336.780 & 57  &  3.1 &1.58& -1.70 \\
Si$\,{\sc ii}$& 6347.091 & SS  &  6.2 &8.12  &  0.15  &   Ti$\,{\sc ii}$& 5381.010 & 42  &  3.3 &1.57& -2.08 \\
Si$\,{\sc ii}$& 6371.359 & SS  &  6.2 &8.12  & -0.08  &   Ti$\,{\sc ii}$& 5418.770 & 33  &  3.1 &1.58& -1.99 \\
Ca$\,{\sc i} $& 4425.444 & 44  &  4.8 &1.88  & -0.36  &   Cr$\,{\sc i} $& 4254.350 &120  &  3.6 &0.00& -0.11 \\
Ca$\,{\sc i} $& 4434.967 & 56  &  4.6 &1.89  & -0.01  &   Cr$\,{\sc i} $& 4274.800 & 73  &  3.2 &0.00& -0.23 \\
Ca$\,{\sc i} $& 4435.688 & 37  &  4.8 &1.89  & -0.52  &   Cr$\,{\sc i} $& 4289.720 & 85  &  3.5 &0.00& -0.36 \\
Ca$\,{\sc i} $& 4454.793 & 90  &  4.7 &1.90  &  0.26  &   Cr$\,{\sc i} $& 5204.510 & 38  &  3.6 &0.94& -0.21 \\
Ca$\,{\sc i} $& 5588.764 & 31  &  4.5 &2.53  &  0.21  &   Cr$\,{\sc i} $& 5206.020 & 60  &  3.6 &0.94&  0.02 \\
Ca$\,{\sc i} $& 6102.722 & 25  &  4.8 &1.88  & -0.79  &   Cr$\,{\sc i} $& 5208.420 & 58  &  3.4 &0.94&  0.16 \\
Ca$\,{\sc i} $& 6122.219 & 37  &  4.6 &1.89  & -0.32  &   Cr$\,{\sc ii}$& 4242.380 & 40  &  3.7 &3.87& -1.33 \\
Ca$\,{\sc i} $& 6162.172 & 57  &  4.6 &1.90  & -0.09  &   Cr$\,{\sc ii}$& 4261.920 & 34  &  3.8 &3.86& -1.53 \\
Sc$\,{\sc ii}$& 4246.840 & 182 &  1.3 &0.32  &  0.24  &   Cr$\,{\sc ii}$& 4555.016 & 20  &  3.5 &4.07& -1.25 \\
Sc$\,{\sc ii}$& 4305.710 & 38  &  1.1 &0.60  & -1.21  &   Cr$\,{\sc ii}$& 4558.660 & 67  &  3.6 &4.07& -0.66 \\
Sc$\,{\sc ii}$& 4314.090 & 160 &  1.5 &0.62  & -0.10  &   Cr$\,{\sc ii}$& 4592.090 & 30  &  3.6 &4.07& -1.22 \\
Sc$\,{\sc ii}$& 4670.404 & 53  &  1.4 &1.36  & -0.58  &   Cr$\,{\sc ii}$& 4616.640 & 40  &  3.9 &4.07& -1.29 \\
Sc$\,{\sc ii}$& 5031.019 & 75  &  1.4 &1.36  & -0.40  &   Cr$\,{\sc ii}$& 4618.820 & 53  &  3.9 &4.07& -1.11 \\
Sc$\,{\sc ii}$& 5239.823 & 33  &  1.3 &1.46  & -0.77  &   Cr$\,{\sc ii}$& 4634.100 & 32  &  3.7 &4.07& -1.24 \\
Sc$\,{\sc ii}$& 5526.809 & 60  &  1.2 &1.77  &  0.02  &   Cr$\,{\sc ii}$& 4824.120 & 66  &  3.9 &3.87& -1.23 \\
Sc$\,{\sc ii}$& 5657.870 & 40  &  1.3 &1.51  & -0.60  &   Cr$\,{\sc ii}$& 4848.240 & 55  &  3.7 &3.86& -1.13 \\
Ti$\,{\sc ii}$& 4367.657 & 63  &  3.1 &2.59  & -0.72  &   Cr$\,{\sc ii}$& 4876.410 & 33  &  3.7 &3.86& -1.47 \\
Ti$\,{\sc ii}$& 4386.850 & 34  &  3.3 &2.60  & -1.26  &   Cr$\,{\sc ii}$& 5237.350 & 32  &  3.6 &4.07& -1.16 \\
Ti$\,{\sc ii}$& 4394.020 & 61  &  3.0 &1.22  & -1.89  &   Cr$\,{\sc ii}$& 5274.990 & 18  &  3.4 &4.07& -1.29 \\
Ti$\,{\sc ii}$& 4395.830 & 74  &  3.5 &1.24  & -2.17  &   Mn$\,{\sc i} $& 4030.753 & SS  &  \leq2.7&0.00&-0.48\\
Ti$\,{\sc ii}$& 4407.680 & 32  &  3.2 &1.22  & -2.47  &   Ni$\,{\sc i} $& 5476.910 & SS  &  4.3 &1.83& -0.89 \\
Ti$\,{\sc ii}$& 4411.100 & 47  &  3.3 &3.10  & -0.62  &   Sr$\,{\sc ii}$& 4077.730 & 172 &  0.2&0.00&  0.14 \\
Ti$\,{\sc ii}$& 4417.720 & 128 &  3.3 &1.17  & -1.43  &   Sr$\,{\sc ii}$& 4215.540 & 159 &  0.2&0.00& -0.18 \\
Ti$\,{\sc ii}$& 4418.310 & 62  &  3.0 &1.24  & -1.82  &   Y $\,{\sc ii}$& 4900.110 & SS  &\leq-0.3&1.03&-0.09 \\
Ti$\,{\sc ii}$& 4441.730 & 28  &  3.1 &1.18  & -2.41  &   Zr$\,{\sc ii}$& 4208.980 & SS  & \leq0.7&0.71&-0.51 \\
Ti$\,{\sc ii}$& 4443.820 & 162 &  3.0 &1.08  & -0.71  &   Ba$\,{\sc ii}$& 5853.675 & 15  &  -0.1 &0.60&-0.91 \\
Ti$\,{\sc ii}$& 4444.540 & 46  &  2.9 &1.12  & -2.03  &   Ba$\,{\sc ii}$& 6141.718 & 75  &   0.0 &0.70&-0.03 \\
Ti$\,{\sc ii}$& 4450.500 & 104 &  2.9 &1.08  &	-1.45 &   Ba$\,{\sc ii}$& 6496.897 & 57  &  0.1  &0.60&-0.41 \\
Ti$\,{\sc ii}$& 4464.460 & 82  &  3.4 &1.16  & -2.08  &   Eu$\,{\sc ii}$& 4205.050 & 30  &  -1.0 &0.00& 0.21 \\
\hline
\hline
        \end{array}
$$
  \end{table*}

 \begin{table*}
    \caption[]{Abundances of the observed species for M79 PAGB star are presented for four different model atmospheres. Also listed are
    abundances of the RGB stars in the same globular cluster analyzed by Carretta et al. (2008, private communication).}
       \label{}
   $$
       \begin{array}{@{}l||c@{}r@{}rrr||@{}c@{}}
          \hline
          \hline
               &            &       	   &  (T_{\rm eff}, \log\,g) &     &     &    \\
\cline{2-6}
 $Species$     &$RGB$       & $(6300,0.80)$ &$(6500,1.18)$  & $(6800,1.67)$   & $(7000,1.98)$ & \log\epsilon_{\odot}^{\star}   \\
\cline{2-6}
   &&  & \log\epsilon(X), $[X/Fe]$ &&   &    \\
          \hline
          \hline
% C$\,{\sc i}$       &  ..., ...  &\leq5.70,\leq-0.68&\leq5.86,\leq-0.66  &\leq5.96,\leq-0.77  &\leq6.06,\leq-0.82  & 8.39   \\
% O$\,{\sc i}$       & 7.29, +0.05&\leq7.70,\leq1.05&\leq8.13,\leq1.34    &\leq8.13,\leq1.13   &\leq8.13,\leq0.98   & 8.66   \\
 C$\,{\sc i}$       &  ..., ...  &\leq5.7,\leq-0.7&\leq5.7,\leq-0.8     &\leq5.9,\leq-0.8     &\leq6.0,\leq-0.9 & 8.39   \\
 O$\,{\sc i}$       & 7.29, +0.05&  8.0,+1.4  & 8.0,+1.2    & 8.0,+1.0     & 7.9,+0.8      & 8.66   \\
% O$\,{\sc i}$^{**}  & 7.29, +0.05&  8.35 ,1.70  & 8.36, 1.57     & 8.36, 1.36	  & 8.44, 1.29     & 8.66   \\
 Na$\,{\sc i}$      & 5.05, +0.42&\leq4.0 ,\leq-0.1&\leq3.8,\leq-0.5 &\leq3.8,\leq-0.7 &\leq3.7,\leq-0.9  & 6.17   \\
% Na$\,{\sc i}$      & 5.05, +0.42&  5.20 ,1.04  & 5.17, 0.87     & 5.37, 0.86	  & 5.60, 0.94     & 6.17   \\
 Mg$\,{\sc i}$      & 6.13, +0.28&  5.8,+0.2  & 5.8,+0.2      & 6.0,+0.1    & 6.1,+0.1      & 7.53   \\
 Mg$\,{\sc ii}$     &  ..., ...  &  5.9,+0.4  & 6.0,+0.4      & 6.0,+0.1    & 6.0,+0.0      & 7.53   \\ 
 Si$\,{\sc ii}$^{\star}& 6.24, +0.29&6.2,+0.7 & 6.2,+0.6      & 6.2,+0.4    & 6.3,+0.3      & 7.51   \\
 Ca$\,{\sc i}$      & 4.93, +0.28&  4.7,+0.4  & 4.8,+0.4      & 4.9,+0.3    & 5.1,+0.3      & 6.31   \\
 Sc$\,{\sc ii}$     & 1.51, +0.04&  1.1,+0.1  & 1.2,+0.1      & 1.5,+0.1    & 1.6,+0.1      & 3.05    \\
 Ti$\,{\sc ii}$     & 3.49, +0.07&  3.2,+0.3  & 3.4,+0.4      & 3.6,+0.4    & 3.8,+0.4      & 4.90   \\
 Cr$\,{\sc i}$      & 3.96, -0.13&  3.5,-0.1  & 3.7,-0.1      & 3.9,-0.1    & 4.1,-0.0      & 5.64   \\
 Cr$\,{\sc ii}$     & 4.14, -0.02&  3.7,+0.1  & 3.8,+0.1      & 4.0,+0.1    & 4.1,-0.0      & 5.64   \\
 Mn$\,{\sc i}$      & 3.23, -0.53&\leq2.7,\leq-0.7&\leq 2.9,\leq-0.6 &\leq 3.1,\leq-0.6  &\leq 3.3,\leq-0.5  & 5.40   \\
 Fe$\,{\sc i}$      & 5.96, +0.00&  5.4,+0.0  & 5.6,+0.0      & 5.8,+0.0    & 5.9,+0.0      & 7.45   \\
 Fe$\,{\sc ii}$     & 5.94, +0.00&  5.4,+0.0  & 5.6,+0.0      & 5.8,+0.0    & 5.9,+0.0      & 7.45   \\
 Ni$\,{\sc i} $     & 4.54, -0.16&  4.3,+0.1  &  4.4,+0.1     & 4.7,+0.1    & 4.8,+0.2      & 6.23   \\
 Sr$\,{\sc ii}$     &  ..., ...  &  0.2,-0.7  & 0.4, -0.6     & 0.8,-0.5    & 1.0,-0.4      & 2.92   \\
 Y $\,{\sc ii}$     & 0.42, -0.27&\leq-0.3,\leq-0.5&\leq -0.1,\leq-0.4 &\leq 0.2,\leq-0.3    &\leq 0.4,\leq-0.3   & 2.21   \\
 Zr$\,{\sc ii}$     & 0.89, -0.16&\leq 0.7,\leq+0.1&\leq 0.9,\leq+0.2 &\leq 1.1,\leq+0.2 &\leq 1.4,\leq+0.4   & 2.59   \\
 Ba$\,{\sc ii}$     & 0.80, +0.13& 0.0,-0.1  & 0.2, -0.1     & 0.5,+0.0     & 0.8,+0.2      & 2.17   \\
 Eu$\,{\sc ii}$     &-0.62, +0.38& -1.0,+0.5  &-0.8, +0.6     &-0.5,+0.7    &-0.2, +0.9     & 0.52   \\
\hline
\hline
       \end{array}
   $$
\begin{list}{}{}
%\vskip 0.2 cm
%\item $^{\mathrm{[N]  }}$ is the number of the lines included in the analysis.$^{\mathrm{[SS]  }}$ Spectrum synthesis. We adopt the
%usual spectroscopic notation: [A]=log(A)star-log(A)$_{\odot}$ for any abundance quantity A; log(A) is the abundance by number of the
%element A in the standard scale where log(H)=12.\\ $[$X/Fe$]$=log(X/X$_{\odot}$)-log(Fe/Fe$_{\odot}$).
%\vskip 0.2 cm 

\item \hskip 1.7 cm \center $^{\mathrm{{^{\star}}}}$ The reported RGB abundance for silicon is from neutral silicon lines.

\end{list}
 \end{table*}

\section{Abundance Analysis -- Elements and Lines}

\noindent For prospective elements, a systematic search was conducted for,
as appropriate, lines of either
the neutral and/or singly-ionized atoms with the likely abundance, lower excitation potential
and $gf$-value as the guides. In this basic step, the venerable
Revised Multiplet Table (Moore 1945) remains a valuable initial
guide.
 When a reference to solar
abundances is  necessary in order to convert our
abundance of element X to either of the quantities [X/H] or [X/Fe],
 we defer to Asplund et al. (2005). In general, many
stellar lines are  strong saturated lines in the solar
spectrum and, therefore, a line-by-line analysis for the stellar-solar
abundance differences is precluded. Carretta provided their adopted
solar abundances which we use to convert their [X/Fe] to abundances
$\log\epsilon$(X). In addition, they provided a list of lines
which were the basis for the line selection made in the analysis of
the red giants. A statistical comparison of the $gf$-values for common
lines suggests that zero-point
differences in abundances arising from different choices of $gf$-value between the RGB stars and the PAGB star are small.
\vskip 0.2 cm
\noindent Our abundances are presented in Table 3 for the consensus model
(6300, 0.80) and the three other models. An error analysis is summarized in
Table 4 where we give the abundance differences resulting from models that
are variously 300 K hotter, +0.2 dex of higher gravity, and experiencing a
$\pm0.5$ km s$^{-1}$ different microturbulence than the consensus model.

\begin{figure}
 \centering
 \includegraphics[width=0.99\columnwidth,angle=0]{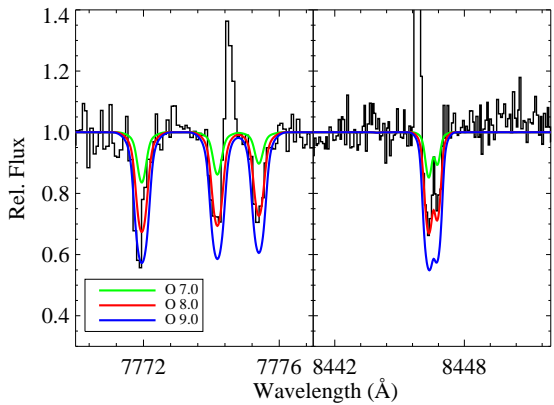}
   \caption{Observed and synthetic spectra around 7774 \AA\  and 8446 \AA\ O\,{\sc i} multiplets. The emission lines
   at 7774.5 and 8446.5 \AA\ are night-sky emission OH lines.}
      \label{f_oxygen_synthesis}
 \end{figure}

\vskip 0.2 cm
\noindent Comments on individual elements follow:
\vskip 0.2 cm
\noindent {\bf C:} Detection of C\,{\sc i} lines  was sought via the 3s$^3$P$^o$-3p$^3$P
multiplet with lines between 9061\AA\ and 9112\AA. The multiplet is absent.
The upper limit to the C abundance for the consensus model is
$\log\epsilon$(C)$\leq 5.7$ from the 9061.44, 9062.49, and 9078.29\AA\ lines
with the $gf$-values from the NIST\footnote{National Institute of Standards and Technology.} database.
\vskip 0.2 cm
\noindent {\bf O: } The O\,{\sc i} triplet at 7774 \AA\ is strong (Figure~\ref{f_oxygen_synthesis});
an abundance $\log\epsilon$(O) $\simeq 8.0$ fits the triplet.
The 8446 \AA\ feature is present and provides
an abundance  $\log\epsilon$(O) $\simeq 8.0$ (Figure~\ref{f_oxygen_synthesis}).  The O\,{\sc i} lines at 9260.8 \AA\
and 9262.7 \AA\ are absent and the upper limit $\log\epsilon$(O) $\leq 8.0$ is
obtained. These consistent abundances correspond to the consensus model and to adoption of the
NIST $gf$-values. Carretta (private communication) has noted that a correction for non-LTE effects
lowers the  7774 \AA\ abundance by about 0.6-0.8 dex. Corrections of a similar magnitude
may apply to the other O\,{\sc i} lines.

\begin{figure}
 \centering
 \includegraphics[width=0.99\columnwidth,angle=0]{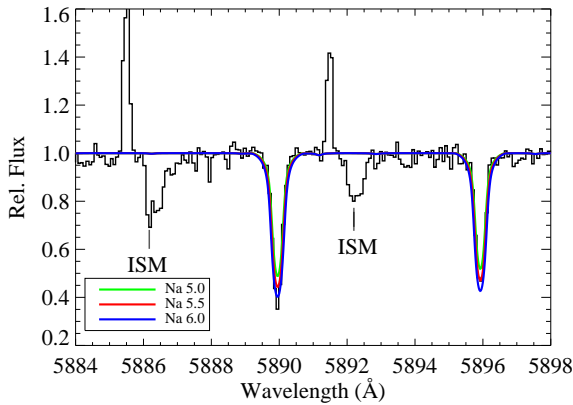}
 \caption{Observed and synthetic spectra around the 5895 \AA\ Na\,{\sc i}  D lines. The interstellar Na D components
are labelled ISM. Night-sky emission components are present. }
 \label{f_na_synthesis}
 \end{figure}

\vskip 0.2 cm
\noindent {\bf Na:} The Na D lines are prominent features in the spectrum (Figure~\ref{f_na_synthesis}).
Three contributors are identifiable: strong stellar D lines, interstellar
D lines with  several blended components, and the night sky emission
features  which are not completely removed in the data reduction. The interstellar
lines are similar as regards strength and velocity with
the interstellar
D lines reported by Gratton \& Ortolani (1989) from spectra of two
cluster red giants.

\noindent The Na abundance $\log\epsilon$(Na) $\simeq 5.5$
provides a fair fit to both D lines with the consensus model.
However, as Figure~\ref{f_na_synthesis} clearly shows,
the stellar Na D lines are ill-suited for an abundance analysis as they
fall on the `flat' part of the curve of growth.
 The Na abundance is
sensitive to small uncertainities in the measured EW and  the adopted
value of the microturbulence.  The EW uncertainty of $\pm$13 m\AA\
translates to an abundance uncertainty of about $\pm$0.2
dex.
A microturbulence uncertainty of $\pm0.5$ km s$^{-1}$ corresponds
to an abundance uncertainty of about $\pm$0.3 dex.
Taking into account the $T_{\rm eff}$
sensitivity of (all) Na\,{\sc i} lines, the likely uncertainty of a Na
abundance derived from the D lines is in the range of $\pm$0.5 dex,
even if LTE were valid.

\noindent In light of the inherent uncertainty in use of the Na D lines,
 we searched for weaker Na\,{\sc i}
lines. The leading candidates are the lines at 8183.3\AA\ and 8194.8\AA.
Neither line is present in our spectrum. Spectrum synthesis gives
the upper limit to the Na abundance as $\log\epsilon$(Na) $\leq 4.0$ for the
consensus model, the value we adopt for the PAGB star.
 A Na abundance at this value provides Na D lines clearly
much weaker than observed. We suppose that the predicted profiles for $\log\epsilon$(Na)=4.0
may be increased to match the observed D profiles by considering
Na absorption from an extended (stationary) atmosphere, non-LTE
effects, and/or an atmospheric structure different from that of
the computed model atmosphere.
\vskip 0.2 cm

\noindent {\bf Mg:} Six Mg\,{\sc i} lines are listed in Table 2 with their
$gf$-values taken from the NIST database. Carretta et al. chose
their $gf$-values from the same source. In calculating
the mean Mg abundance from Mg\,{\sc i} lines we drop the two strong lines
(5183.6 \AA\ and 8806 \AA) as their derived abundances are sensitive
to the adopted microturbulence and EW uncertainities.

 \begin{figure}
 \centering
 \includegraphics[width=0.99\columnwidth,angle=0]{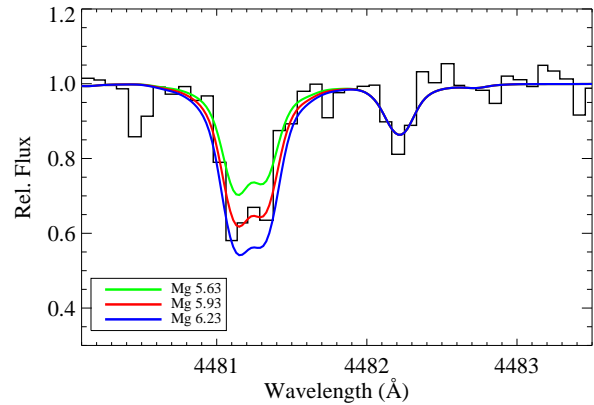}
   \caption{Observed and synthetic spectra around the 4481 \AA\ Mg\,{\sc ii} triplet lines.}
      \label{f_mg_synthesis}
 \end{figure}

\noindent For the Mg\,{\sc ii} 4481\AA\ feature, we take the $gf$-values also
from the NIST database. Spectrum synthesis is used to provide the
Mg abundance - see Figure~\ref{f_mg_synthesis}. For the consensus model, the Mg
abundances from the Mg\,{\sc i} and Mg\,{\sc ii} are 5.8 and 5.9,
respectively.

\begin{figure}
 \centering
 \includegraphics[width=0.99\columnwidth,angle=0]{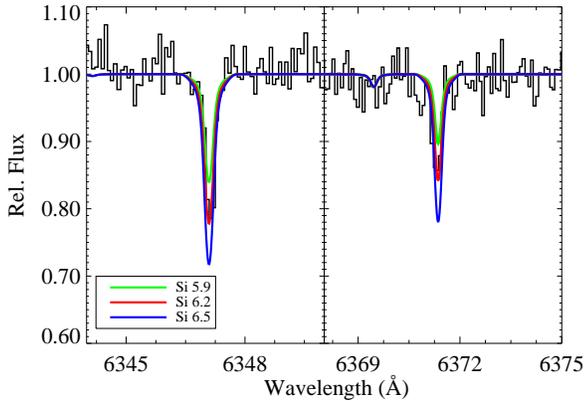}
   \caption{Observed and synthetic spectra around the 6347 and 6371 \AA\ Si\,{\sc ii} lines.}
      \label{f_si_synthesis}
 \end{figure}

\vskip 0.2 cm
\noindent {\bf Si:}  Silicon is represented solely by the Si\,{\sc ii}  lines at
6347 \AA\ and 6371 \AA. For the Si\,{\sc ii} doublet, the $gf$-values are those
recommended by Kelleher \& Podobedova (2008, see also the NIST database).
Spectrum syntheses for three Si abundances are shown with the
observed spectrum in Figure~\ref{f_si_synthesis}. For the consensus model, the syntheses
show that the abundance $\log\epsilon$(Si)=6.2 is a good fit to both
lines with a poorer fit occurring for abundances differing by more than
about $\pm$0.15  dex from this value.

\noindent A search for Si\,{\sc i} lines proved
negative. The tightest limit on the Si abundance from Si\,{\sc i} lines is obtained from
a line at 7282.8 \AA\ with the $gf$-value taken from Lambert \& Warner (1968).
This upper limit to the Si abundance is clearly in excess of that
from the Si\,{\sc ii} lines and the condition of ionization equilibrium, i.e.,
the constraint on the Si locus in Figure~\ref{f_logg_teff} is compatible with the loci for
Mg, Cr, and Fe.
\vskip 0.2 cm
\noindent {\bf Ca:} The Ca abundance is based on the Ca\,{\sc i} lines listed in
Table 2 with their $gf$-values drawn from the NIST database. Carretta's
$gf$-values include NIST values but also Smith \& Raggett's (1981)
measurements for some lines. The difference between  our and Carretta's
$gf$-values varies from line to line but on average is probably
small (say, $< 0.05$ dex) and dependent on the particular line
selection made by Carretta. 

%\vskip 0.2 cm
%\noindent The observed spread in Ca abundance is real and  mainly caused by 4454 \AA\, and 6102 \AA\ Ca\,{\sc i}
%lines. The 6102 \AA\ line has the highest deviation ($\approx$0.4 dex) from the mean. It is followed by 4454 \AA\ by
%$\approx$0.3 dex. In the study of alkaline earths Mg, Ca, Sr, and Ba by Lambert \& Warner (1968), the discordant
%results for above lines were reported to be unknown. They did not include these two lines in the calculation of the
%mean calcium abundance from their Table 5. The reason for relatively high line-to-line scatter for 4454 \AA\, and
%6102 \AA\ is unknown.

\begin{figure}
 \centering
 \includegraphics[width=0.99\columnwidth,angle=0]{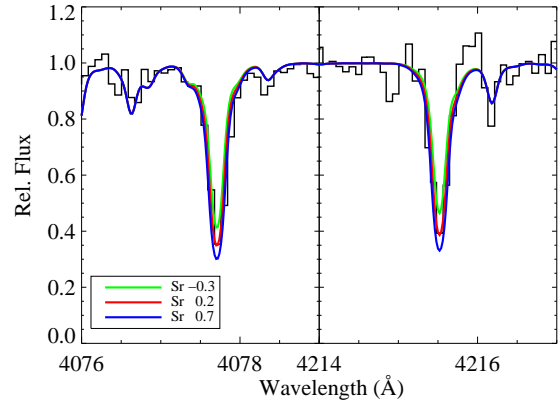}
   \caption{The observed spectrum around the Sr\,{\sc ii} 4077.7 \AA\ and 4215.5 \AA\  resonance lines.}
      \label{f_sr_synthesis}
 \end{figure}

\vskip 0.2 cm
\noindent {\bf Sc:} Our $gf$-values are taken with one exception from the
NIST database. The exception is for the line at 4305\AA\ which is in the
NIST database but without an entry for the $gf$-value. For this line
we adopt the value given by Gratton et al. (2003); Carretta
adopts Gratton et al.'s recommendations for all Sc\,{\sc ii}
lines. With half weight given to the strongest lines (4246\AA\ and 4314\AA)
and the 4305\AA\ line with the non-NIST $gf$-value, the mean Sc
abundance for the consensus and other models is given in Table 3. These
mean abundances are approximately 0.1 dex greater than the value
based on Gratton et al.'s $gf$-values.
\vskip 0.2 cm
\noindent {\bf Ti:} The search for Ti\,{\sc i} lines was unsuccessful. The spectrum of
Ti\,{\sc ii} is well represented.  For the ion, we take the $gf$-values from the NIST
database. The exceptions are Ti\,{\sc ii} lines at 4367, 4394, 4411, and 4418 \AA. For
4394 \AA\ line, we use $gf$-value from Roberts et al. (1973) and for the rest of four,
$gf$-values from Ryabchikova et al. (1994) have been adopted. The abundances from NIST
$gf$-values for these four differ from the mean of about $\pm0.5$ dex. Carretta's line
list is also based on the NIST database.  If a few outliers returning a high abundance are
neglected, the Ti abundance from the Ti\,{\sc ii} lines is independent of the EW of a line. The abundance for the consensus model is
$\log\epsilon$(Ti) = 3.2.
\vskip 0.2 cm
\noindent {\bf Cr:} The Cr\,{\sc i} lines in Table 2  have the well
determined laboratory $gf$-values (Blackwell, Menon \&
Petford 1984; Tozzi, Brunner \& Huber 1985) adopted for the
NIST database and confirmed recently by Sobeck et al. (2007).
Nilsson et al. (2006) from measurements of radiative
lifetimes and branching ratios provide $gf$-values for many
Cr\,{\sc ii} lines, all with wavelengths short of 4850\AA.
The NIST database includes $gf$-values for all but three of
our lines; Carretta adopts the NIST database.
 For six common lines, the mean abundance
from Nilsson et al. is 0.01 dex smaller than from the NIST
compilation.
The dispersion from this subset of
Cr\,{\sc ii} lines is the same for both sets of
$gf$-values.
The Cr abundance from the Cr\,{\sc i} lines is 3.5 and from the
Cr\,{\sc ii} lines is 3.7 dex from the consensus model.
\vskip 0.2 cm
\noindent {\bf Mn:} Lines of Mn are not detectable in our spectrum. The
upper limit for the Mn abundance is best determined from the
absence of the Mn\,{\sc i} resonance lines near 4030\AA\
with their $gf$-values known reliably from laboratory measurements for
which we adopt the NIST database's recommendations.
For the consensus model, the upper limit is $\log\epsilon$(Mn) $\leq 2.7$.
Carretta
adopt NIST results for all Mn\,{\sc i} lines in the spectra of
the red giants. We assume that the RGB Mn abundance may be
compared directly with the upper limit here determined.
\vskip 0.2 cm
\noindent {\bf Fe:} The $gf$-values for Fe\,{\sc i} and Fe\,{\sc ii} lines
are from Fuhr \& Wiese (2006). Carretta's analysis of RGB stars uses $gf$-values
from Gratton et al. (2003) that
are essentially those of Fuhr \& Wiese. Abundance analysis of our lines using the
gf-value selection  made by Gratton et al. gives an abundance only 0.06 dex higher
from Fe\,{\sc i} lines and 0.09 dex higher from Fe\,{\sc ii} lines than
our result. These differences indicate that the PAGB and RGB Fe abundances
are essentially   on the
same scale.
\vskip 0.2 cm
\noindent {\bf Ni:} One Ni\,{\sc i} line is listed in Table 2 with its
$gf$-value from the NIST database. The Ni abundance for the
consensus model is $\log\epsilon$(Ni) = 4.3.
Carretta's list of Ni\,{\sc i} lines does not include this line
but Carretta's $gf$-values are essentially on
the NIST scale.
\vskip 0.2 cm
\noindent {\bf Sr:} The Sr\,{\sc ii} resonance lines at 4077\AA\ and 4215\AA\ are
strong lines: Figure~\ref{f_sr_synthesis} shows these lines together with synthetic
spectra. The $gf$-values are from Brage et al. (1998). These lines
fall near the damping portion of the curve of growth and the
derived abundance is, therefore, sensitive to the small uncertainties in the
EWs, the microturbulence and to the damping (radiative) constant. The
insensitivity of the predicted line profile to the Sr abundance is well illustrated by the
synthetic spectra for Sr abundances of $-0.3$, $+0.2$, and $+0.7$.

\vskip 0.2 cm
\noindent {\bf Y:} In our search for Y\,{\sc ii} lines, Hannaford et al.'s (1982)
list of clean solar Y\,{\sc ii} lines with accurate $gf$-values
was our starting point. Seven solar lines longward of 4300\AA\
and with a solar EW greater than 35m\AA\ were considered.  A
representative line is included in Table 2: a series of synthetic spectra
for
this line at 4900.12\AA\  suggest an upper limit
 $\log\epsilon$(Y) $\leq -0.3$ for the
consensus model.
\vskip 0.2 cm
\noindent {\bf Zr:} Our search for Zr\,{\sc ii} lines drew on the papers by
Bi\'{e}mont et al. (1981) and Ljung et al. (2006) who  measured
accurate laboratory $gf$-values and conducted an analysis of
Zr\,{\sc ii} lines to determine the solar Zr abundance. The
strongest solar line identified with Zr\,{\sc ii} is at
4208.98 \AA.
Synthetic spectra provide the upper limit  $\log\epsilon$(Zr) $\leq 0.7$
for the consensus model.

\vskip 0.2 cm
\noindent {\bf Ba:}  The Ba\,{\sc ii} resonance lines at 4554 \AA\ and 4934 \AA\ are present in
great strength. The abundance analysis is based on the triplet of excited lines from the
5d$^2$D levels at 5853 \AA, 6141 \AA, and 6496 \AA\ with $gf$-values taken as the mean of the
experimental values from Gallagher (1967) and Davidson et al. (1992). The lines have EWs
such that consideration of hyperfine and isotopic splitting may be neglected. (The NIST
$gf$-values for the resonance and these excited lines are erroneous. In compiling these values, it
would appear that the absorption $f$-value given by Davidson et al. was treated as a $gf$-value, i.e., the NIST entries should be multiplied by the statistical weight $g$ of the
lower level of the transition to obtain the $gf$-value.)
\vskip 0.2 cm
\noindent {\bf Eu:} The Eu\,{\sc ii} resonance line at
4205.05\AA\ is present.The weaker line at
4129.73 \AA\ may be present.  The $gf$-values
are the accurate values provided by Lawler et al. (2001).
The abundance is $\log\epsilon$(Eu) = $-1.0$ for the
consensus model from the 4205 \AA\ line with the 4129 \AA\ line providing
weak confirmation.
\vskip 0.2 cm
\noindent The formal errors for the abundances arising from uncertainties
of the atmospheric parameters -- $T_{\rm eff}, \log$ g, and $\xi$ --
are summarized in Table 4  for changes with respect to the
consensus model of $+300$K, $+0.2$ cgs units, and $\pm0.5$ km s$^{-1}$
for the selected lines (i.e., the Na entries are based on the
EW upper limits for the 8183 \AA\ and 8194 \AA\ lines). Given the
estimated combined uncertainties and the dispersion in the
line-to-line abundances for a given element, we elect to
quote the abundances in Table 3 to one decimal place.

%%%%%%%%%%%%%%%%%%%%%%%%%%%%%%%%%%%%%%%%%%%%%
%%% ERRORS in the model atmosphere parameters
%%%%%%%%%%%%%%%%%%%%%%%%%%%%%%%%%%%%%%%%%%%%%

%SECOND TABLE FOR 6300,1.0 consensus model

 \begin{table}
    \caption[]{Sensitivity of the derived abundances to the uncertainties in the model atmosphere parameters for the consensus model.}
       \label{}
   $$
       \begin{array}{@{}l||rrcc}
          \hline
          \hline
         &                 & \Delta log\,\epsilon&     &             \\
\cline{2-5}
$Species$         &  \Delta T_{\rm eff} & \Delta log\,g & \Delta\,\xi  &\Delta\,\xi   \\
\cline{2-5}

  &    $+300$       &  $+0.2$      &  $+0.5$     & $-0.5$ \\
  &       ($K$)     &   ($cgs$)    & ($km$\,s^{-1})& \\
          \hline
C$\,{\sc i}$      &-0.04&-0.05&\hspace*{7pt}0.00&0.00 \\
O$\,{\sc i}$      &-0.15&-0.09&-0.05&0.06  \\
Na$\,{\sc i}$     &+0.20& 0.00&\hspace*{7pt}0.00&0.00  \\
Mg$\,{\sc i}$     &+0.17&-0.01&-0.02&0.02  \\
Mg$\,{\sc ii}$    &-0.10&+0.10&-0.05&0.15  \\
Si$\,{\sc ii}$    &-0.10&+0.05&-0.02&0.05  \\
Ca$\,{\sc i}$     &+0.22&-0.03&-0.03&0.03  \\
Sc$\,{\sc ii}$    &+0.11&+0.05&-0.05&0.05  \\
Ti$\,{\sc ii}$    &+0.12&+0.06&-0.07&0.10  \\
Cr$\,{\sc i}$     &+0.28&-0.01&-0.04&0.07  \\
Cr$\,{\sc ii}$    &+0.05&+0.06&-0.01&0.04  \\
Mn$\,{\sc i}$     &+0.35&-0.05&\hspace*{7pt}0.00&0.00  \\
Fe$\,{\sc i}$     &+0.25&-0.01&-0.03&0.04  \\
Fe$\,{\sc ii}$    &+0.08&+0.06&-0.05&0.08  \\
Ni$\,{\sc i}$     &+0.20& 0.00&\hspace*{7pt}0.00&0.00  \\
Sr$\,{\sc ii}$    &+0.24&+0.05&-0.27&0.39  \\
Y$\,{\sc ii}$     &+0.15&+0.05&\hspace*{7pt}0.00&0.00  \\
Zr$\,{\sc ii}$    &+0.15& 0.00&\hspace*{7pt}0.00&0.00  \\
Ba$\,{\sc ii}$    &+0.30&+0.03&-0.02&0.04  \\
Eu$\,{\sc ii}$    &+0.23&+0.04&-0.02&0.01  \\
          \hline
          \hline
%C$\,{\sc i}$      &+0.21&+0.11& 0.00&0.33  \\  ; 0.01 dex duyarlilik 
%O$\,{\sc i}$      &-0.15&+0.10& 0.00&0.29  \\
%Na$\,{\sc i}$     &+0.25&-0.05&-0.05&0.34  \\
%Mg$\,{\sc i}$     &+0.16&-0.02&-0.01&0.25  \\
%Mg$\,{\sc ii}$    &-0.10&+0.10& 0.00&0.26  \\
%Si$\,{\sc ii}$    &-0.10&+0.10& 0.00&0.26  \\
%Ca$\,{\sc i}$     &+0.22&-0.02&-0.01&0.29  \\
%Sc$\,{\sc ii}$    &+0.11&+0.05&-0.03&0.25  \\
%Ti$\,{\sc ii}$    &+0.11&+0.06&-0.05&0.27  \\
%Cr$\,{\sc i}$     &+0.28&-0.02&-0.03&0.33  \\
%Cr$\,{\sc ii}$    &+0.05&+0.06&-0.01&0.20  \\
%Mn$\,{\sc i}$     &+0.24&-0.06& 0.00&0.32  \\
%Fe$\,{\sc i}$     &+0.25&-0.02&-0.01&0.31  \\
%Fe$\,{\sc ii}$    &+0.07&+0.07&-0.03&0.24  \\
%Ni$\,{\sc i}$     &+0.25& 0.00& 0.00&0.29  \\
%Sr$\,{\sc ii}$    &+0.22&+0.05&-0.17&0.38  \\
%Y$\,{\sc ii}$     &+0.09&-0.06&-0.06&0.26  \\
%Zr$\,{\sc ii}$    &+0.15&+0.05& 0.00&0.26  \\
%Ba$\,{\sc ii}$    &+0.29&+0.03&-0.01&0.33  \\
%Eu$\,{\sc ii}$    &+0.23&+0.04&-0.01&0.31  \\
       \end{array}
   $$
%\begin{list}{}{}
%\item \hskip 1.0 cm $^{\mathrm{{^{\dagger}}}}$ $\sqrt{\sigma^{2}} = \sqrt{(\Delta T_{\rm eff})^2+(\Delta
%log\,g)^2+(\Delta \xi)^2}$
%\end{list}
 \end{table}

\section{DISCUSSION}

This exploration through quantitative spectroscopy of the newly discovered
PAGB star in the globular cluster M79 has led to an unexpected and, therefore,
fascinating result: the standard LTE analysis of the star has resulted in
a metallicity different from that of the RGB stars analysed also by
standard LTE techniques by Carretta.  The consensus model of (6300,0.8) provides
a [Fe/H] of $-2.0$ but the RGB analysis gives a [Fe/H]  of $-1.5$.
Here, we open a discussion of
how one might account for this unexpected difference which exceeds a possible
difference that might arise from uncertainties in the abundance analysis
of these two different kinds of stars.
\vskip 0.2 cm
\noindent Application of photometric and spectroscopic indicators of the
atmospheric parameters for the PAGB star led to the consensus choice of
$T_{\rm eff}=6300$ K and $\log g$=0.8. A model with these parameters
(and a microturbulence $\xi=3.4$ km\,s$^{-1}$) fits not only the indicators but also the locus in the $T_{\rm eff}$ versus $\log g$
plane provided by the constraint on the star's luminosity and
mass. Table 3 summarizes the PAGB star's composition for such a model
atmosphere and contrasts it with the mean composition of the RGB stars. The PAGB star's Fe
abundance is $-0.5$ dex lower than that
of the RGB stars. For the majority of the investigated elements, the
difference in abundance $\log\epsilon$(X) in the sense (Ours $-$ Carretta)
is within the range $-0.5\pm0.3$ dex, i.e., the differences are
equal to $-0.5$ dex to within measurement uncertainties.
This result is readily seen from Table 3 by looking at the entries for [X/Fe]
for elements measured in both the RGB and PAGB stars;
 The exceptions are
O, Na, and Si.  Strontium should be added to this trio because one
anticipates that [Sr/Fe] $\simeq$ [Y/Fe] $\simeq$ [Zr/Fe]  but [Sr/Fe] for the
PAGB star is apparently 0.7 dex less than [Y/Fe] and [Zr/Fe] for the
RGB stars.
\vskip 0.2 cm
\noindent The exceptions of O and Na may be related. Carretta notes that
the RGB stars show the O-Na correlation found in other
globular clusters: high-Na paired with low-O. Here, the PAGB star
has ([O/Fe],[Na/Fe]) = ($1.4,<-0.1$) but the mean for the 10 RGB stars
is ($+0.05,+0.42$). A likely contributor to the explanation, as suggested by its lower Na
abundance is that the PAGB star
evolved from an RGB star largely uncontaminated with the products
of the H-burning reactions that consume O and produce Na. The [O/Fe] of $+1.4$ is
higher than expected for unevolved metal-poor stars. Estimated non-LTE effects on the
O\,{\sc i} 7770 \AA\ triplet lines reduce the O abundance by about 0.6 -- 0.8 dex. If
similar corrections apply to the 8446 \AA\ feature, the [O/Fe] would be cut from +1.0 to
+0.2 -- 0.4, an expected value for a metal-poor star as the value reported for the
cluster's RGB stars. (Carretta et al. used the [O\,{\sc i}]
lines that are unaffected by non-LTE effects.)
\vskip 0.2 cm
\noindent The [X/Fe] for the PAGB star  are generally consistent with
expectation for metal-poor stars and, as noted above, generally
equal to the measured values for the cluster's RGB stars. The limit on
[C/Fe] ($< -0.8$) indicates reduction of C by the first dredge-up and
the absence of a C enrichment on the AGB.  The [X/Fe] for the $\alpha$-elements
(Mg, Ca, and Ti) at about $+0.3$ show the characteristic  enhancement
for metal-poor stars. Silicon is a notable exception -- [Si/Fe] $= +0.7$ --
for the PAGB star. The RGB stars have the expected value for [Si/Fe]  of $+0.3$. This
difference may be due to the use of the Si\,{\sc ii} lines for the
PAGB star and Si\,{\sc i} lines for the RGB stars; non-LTE effects may
have enhanced the strengths of the Si\,{\sc ii} lines. Strontium was not
measured in the RGB  stars but our result [Sr/Fe] $= -0.7$ is, as noted above, strongly at
odds with the RGB stars' values of $-0.3$ for Y and $-0.2$ for Zr.
 This
discrepancy is discussed below.
\vskip 0.2 cm
\noindent Are these differences between the PAGB star and the RGB stars
 in [X/H] for many elements and [X/Fe] for a
few elements
intrinsic differences
or  reflections of systematic
errors in the analyses?
Certainly, the differences cannot be eliminated by an alternative choice
of atmospheric parameters
within the confluence of the
various loci in Figure~\ref{f_logg_teff}, i.e., models with $T_{\rm eff}$ within 300 K
and $\log g$ within 0.3 dex of the consensus model (see Table 4).
 These questions we discuss next.
\vskip 0.2 cm
\noindent Cluster RV Tauri variables analysed previously have not
shown a clear [Fe/H] difference with the [Fe/H] of the RGB stars of the host
cluster. Gonzalez \& Lambert (1997) analysed variables in
M2, M5, M10, and M28. With the exception of one variable in
M5, the $T_{\rm eff}$  (spectroscopically derived from Fe\,{\sc i} lines)
were cooler than 5750 K. For three stars in clusters with [Fe/H] from
$-1.2$ to $-1.6$, the RV Tauri star's [Fe/H] was found to be within 0.1 dex of that
from the RGB stars. The exception was a cluster (M10) with the RV Tauri
star at $T_{\rm eff}=4750$ K giving [Fe/H]$=-2$ and  with RGB stars showing
[Fe/H]$=-1.5$ -  an intriguing
parallel with M79? The hottest variable in the sample (M5 V42, a star remarkable
for the presence of Li) was reanalysed
by Carney et al. (1998) using three spectra providing  spectroscopic
temperatures of 5200, 5500, and 6000 K. The [Fe/H] was $-1.2$ from each
spectrum and equal to that from the cluster's RGB stars.
It is surely ironic that adoption of plane-parallel atmospheres for
the analysis of these variable stars returns
the same composition as their cluster's RGB but the use of the model atmospheres
for the (apparently) non-variable PAGB star in M79 gives rise to the abundance
difference with the RGB stars.

\subsection{Evolution and abundances}

In the evolution of a RGB star to a PAGB star, there are
several processes that may affect the surface
compositions. At the tip of the RGB, the He-core flash and
attendant mass loss
may mix products of He burning to the
surface. Evolution along the AGB following He-core burning may
through the third dredge-up bring C and $s$-process products to the
surface. Severe mass loss along and at the tip of the AGB may
enhance changes in surface composition.
\vskip 0.2 cm
\noindent Additionally, the surface composition of a PAGB star may be
altered by processes unconnected to internal nucleosynthesis and
dredge-up. PAGB stars have been found with abundance anomalies
correlated with (i) the condensation temperature at which an
element condenses out as dust or onto dust, and (ii) the ionization
potential of the neutral atom, the so-called first ionization potential or
FIP effect.
Dust-gas separation, also referred to as winnowing, is possibly associated with a circumbinary
dusty disk from which gas but not dust is accreted by the
AGB and/or PAGB star (Van Winckel 2003).
The signature of dust-gas winnowing is an underabundance correlated with
the element's predicted condensation temperature ($T_{\rm C}$).
 Observational examination of the
dust-gas winnowing phenomena suggests that it is ineffective in
intrinsically metal-poor stars, say [Fe/H] $<-1$ (Giridhar et al.
2005), and, therefore, is unlikely to have been effective in
the M79 star, even were it a binary. Indeed, the RV Tauri stars in
M2, M5, M10, and M28 (see Introduction) do not show evidence of
dust-gas winnowing.
Rao \& Reddy (2005) show that  field RV Tauri
variables not portraying a clear signal of dust-gas winnowing
 may show abundance anomalies correlated with the ionization
potential of the neutral atom (the FIP
effect).
Our search (see below) for an explanation of the composition difference between the
PAGB and RGB stars in terms of  nucleosynthesis and dredge-up,
 dust-gas winnowing, and   the FIP effect proves
negative.

\subsubsection{Nucleosynthesis and dredge-up}

The PAGB star's C abundance is evidence that the PAGB star's progenitor either
did not  experience  C-enrichment from the third
dredge-up or the enrichment was subsequently erased by (presumably) H-burning.
 The upper limit [C/Fe] $< -0.7$ is consistent with
 observations of other
globular clusters that  give [C/Fe] $<0$ with a decrease
along the RGB, as anticipated from the first dredge-up.
 (Carretta does not report a
C abundance for the  RGB stars in M79.)
Apparently, the He-core flash at the tip of the RGB and evolution from the
horizontal branch to the AGB did not result in C enrichment from the
addition of He-burning products to the atmosphere.
\vskip 0.2 cm
\noindent  Comparison of the heavy element abundances for the PAGB star with
results for the RGB stars appears to provide a fascinating puzzle.
For the RGB stars, Carretta reports
 [X/Fe] = $-0.27, -0.16, +0.13$ for Y, Zr, and Ba ($s$-process
indicators) and $+0.38$ for Eu  ($r$-process indicator), with [Fe/H]
$=-1.6$. The star-to-star intrinsic variation in these [X/Fe] is
less than the small
observational uncertainties. These mean  values are  not at all unusual;
field stars
 and stars in
globular clusters of a comparable  metallicity have these values (Gratton et al. 2004).
For the PAGB star, our results are [X/Fe] = $< -0.5$, $<+0.1$ and $-0.1$ for
Y, Zr, and Ba and $+0.5$ for Eu for the consensus model. The upper limits for
Y and Zr are consistent with the values for the RGB stars.
The [Ba/Fe]  is less than for
the RGB stars but possibly within the range of observational
uncertainties considering that non-LTE effects have not been considered for
either the PAGB star or the RGB stars. In short, there is no evidence that the
PAGB star is a victim of $s$-processing on the AGB.
Our [Eu/Fe]$\simeq+0.5$ matches the RGB stars' Eu abundance.
The outstanding puzzle is the Sr abundance of the PAGB star.
\vskip 0.2 cm
\noindent
The [Sr/Fe] $= -0.7$ is clearly
at odds with [Y/Fe]  $\simeq$ [Zr/Fe] $\simeq -0.2$ for the RGB stars.
Moreover, among metal-poor stars -- see the
sample displayed by  Lambert \& Allende Prieto (2002) -- the ratio [Sr/Ba] is positive
but here [Sr/Ba] is highly negative ($-0.6$); the PAGB star would then be a remarkable outlier far from
the representative sample shown by Lambert \& Allende Prieto.
It may be possible to conceive of a $s$-process scenario
that  reduces all three of these ratios yet keeps  [Ba/Fe] close to
the value for the RGB stars, but
even highly contrived operating conditions for the $s$-process may not
exist for producing the low [Sr/Fe] relative to [Y/Fe] and [Zr/Fe].
\vskip 0.2 cm
\noindent The key to the Sr problem may be that the Sr\,{\sc ii} lines
are strong and thus the Sr abundances are particularly uncertain with the
microturbulence and
non-LTE effects as major contributors to uncertainty.
 The consensus model adopts $\xi = 3.4\pm0.6$ km s$^{-1}$.
If microturbulence is reduced to about $\xi = 2.8$ km s$^{-1}$, the
Sr abundance is raised to [Sr/H] $\simeq 0.7$ and [Sr/Fe] $\simeq -0.2$,
a value consistent with the results for Y and Zr among the RGB stars. This value of $\xi$ is certainly not excluded by Figure 5. If one
were confident that other effects (i.e., non-LTE) were not
affecting the Sr\,{\sc ii} lines in a serious way, one might
use the Sr abundance to set $\xi$ for the abundance analysis. This
alternative choice for $\xi$ has a minor effect on the other abundances
except for the few elements represented exclusively by strong lines.
The Sr\,{\sc ii} line profiles may be fitted with this lower $\xi$ and
a modest value for the macroturbulence. The key to resolving the
Sr question will be observations of the 4d$^2$D - 5p$^2$P$^o$
multiplet with lines at 10036.6 \AA, 10327.3 \AA, and 10914.9 \AA.

\subsubsection{Dust-gas winnowing and the FIP effect}

A search for a correlation with the condensation temperature ($T_{\rm C}$) for
an element returns a negative result. Some field RV Tauri
variables show a striking correlation, e.g., HP Lyr (Giridhar
et al. 2005). In HP Lyr, the Sc, and Ti underabundances at
$T_{\rm C} \simeq 1600$ K are 2 dex
greater than for Fe with $T_{\rm C}= 1330$ K,
 Ca with $T_{\rm C}= 1520$ K is 1 dex more underabundant,
 and Na at $T_{\rm C}=960$ K is 1 dex
more abundant than Fe. For the consensus  model, the [X/Fe] are here not
correlated at all with $T_{\rm C}$.
 There is no indication that Sc and Ti in the PAGB star are underabundant
relative to Fe and other elements of similar $T_{\rm C}$.
Indeed, as noted above, the
[X/Fe] of the PAGB star are not only generally equal to the values for the
RGB star where dust-gas winnowing may be dismissed as a possible
influence, but also essentially equal to the [X/Fe] expected for a
metal-poor star.
The Sr anomaly identified above is not attributable to dust-gas winnowing;
the $T_{\rm C}$ of Sr is intermediate between that of Ca and Fe.
Similarly there is no correlation of the abundances with the FIP.

\subsection{Alternative choices of model atmosphere}

If matching the [Fe/H] of the PAGB star to that of the RGBs
is an overriding requirement in the selection of the model
atmosphere parameters along with retention of the assumption of LTE in the
model's construction and the abundance
analysis, one may pursue $T_{\rm eff},\log g$ combinations
that lie along the line of Fe ionization equilibrium. A combination such as (7000, 1.98) is
needed. Abundances for this and intermediate combinations are
given in Table 3.
\vskip 0.2 cm
\noindent This approach clearly provides Balmer line profiles that are
far broader than the observed profiles (Figure~\ref{f_balmer_profiles}), an
Fe abundance from Fe\,{\sc i} lines that decreases with increasing
excitation potential (Figure~\ref{f_excitation}).
There is yet another problem raised by
use of the (7000,1.98) model: this combination lies far off the
locus set by the $L-M$ relation (Figure~\ref{f_logg_teff}). To increase the
$\log g$ from this relation requires an increase by 0.85 dex in the
ratio $\log M/L$ (equation 1). Clearly, the allowable range on $M$ does not
permit anything but a very small part of this increase. The major
part has to be attributed to a decrease of the luminosity $L$. If one
entertains the not implausible idea that the PAGB star may be affected
by circumstellar dust, one immediately realises that
extinction from a circumstellar dust shell aggravates the problem
because the $L$ inferred from the magnitude $m_V$ is then increased not
decreased. Luminosity $L$ may be decreased by a substantial reduction of
the distance modulus but this negates the association of the
star with M79.
In short,
matching the PAGB star's Fe abundance to that of the RGB stars by
choosing a much hotter classical model for the PAGB star
that is consistent with the ionization
equilibrium of Fe (and approximately Mg and Cr too) does not result in a satisfactory
understanding of the PAGB star.
\vskip 0.2 cm
\noindent This unsatisfactory state of affairs may in part arise because of the
retention of the assumption of LTE.  To first-order the principal
non-LTE effect may be the over-ionization of Fe and similar metals.
This over-ionization has a major effect on the strength of the
lines of Fe\,{\sc i} and similar metals. Since the Fe is
predominantly present as Fe$^+$, such non-LTE effects are weak
for Fe\,{\sc ii} lines. Also, effects of the overionization on the
structure of the atmosphere are anticipated to be  slight too because the
additional contribution of electrons attributable to this
non-LTE effect is very small.  Thus, one might suppose that the
Fe\,{\sc ii} lines are a safe indicator of the PAGB star's Fe
abundance.

\vskip 0.2 cm
\noindent Discarding LTE in this way and raising the PAGB star's Fe abundance from
the Fe\,{\sc ii} lines to
the RGB stars' value (where we assume non-LTE effects are minimal
for the giants) requires   a combination of a higher
surface gravity and a higher  effective temperature. The desired 0.6 dex increase in [Fe/H] is very difficult to
achieve by sliding slightly the PAGB star along the
$L-M$ relation; Table 4 shows that $T_{\rm eff}$ and $\log$g increases both raise the
Fe abundance derived from the Fe\,{\sc ii} lines. A very extreme slide to
$T_{\rm eff}=7500$ K and $\log g=1.5$ increases the
Fe abundance from Fe\,{\sc ii} lines by the desired 0.5 dex.
Thus, our simple assessment of likely non-LTE
effects does not ease at all the reconciliation of the
PAGB and RGB Fe abundances.
\vskip 0.2 cm
\noindent One is seemingly forced to the conclusion that the model
atmosphere may require substantial revision for the PAGB
star. One concern is that the model atmospheres and the line analysis
program assume a plane-parallel geometry but a spherical
geometry may be more appropriate for  the PAGB star. Heiter \&
Eriksson (2006) evaluated this concern for stars of solar
metallicity and temperatures and gravities spanning the values of our
PAGB star. Their conclusion is that abundance errors resulting from
use of the plane-parallel geometry in model and line analysis rather than
a consistent use of spherical geometry are minor provided that the
analysis is restricted to weak (EW $<$ 100 m\AA) lines.
Heiter \& Eriksson express the expectation that this conclusion is
likely valid too for low-metallicity stars and this was verified
by Eriksson (2009, private communication) for this PAGB star.
Apart from a few
elements (e.g., Sr), our analysis is indeed based primarily on
weak lines.
Thus, the generally lower abundances for the PAGB star
relative to the RGB  stars are not attributable to use of the
inappropriate geometry. An alternative representation of the
model atmospheres seems necessary. Reproduction of the observed line strengths with the RGB
star's composition may be possible by invoking
a departure from the structure of
line-forming regions predicted by the
ATLAS9 model. Two obvious qualitative possibilities are the
presence of stellar granulation on small and/or large scales
and a flatter temperature gradient over a uniform atmosphere.
Such a change in gradient requires a different
balancing of heating and cooling in the affected regions. Perhaps,
convective transport of energy is inadequately modelled and/or
`mechanical energy' (e.g., acoustic waves) is supplied from the
deep convective envelope. Of course, a revised atmospheric structure
invalidates the determination of atmospheric parameters based on ATLAS9
models. Detailed exploration of semi-empirical atmospheres with
might be undertaken to see how or if an
atmospheric structure may be discovered that reconciles the
wide variety of line strengths from the many elements with the
composition of the RGB stars.

This suggestion about a failure of classical atmospheres to predict
correctly the temperature gradient
echoes a previous similar suggestion concerning the atmospheres of the
R CrB stars, warm supergiants  with a very H-deficient composition
(Asplund et al. 2000). For those stars, the diagnostic indicating  the
failure was the strengths of the C\,{\sc i} lines. The strengths of these
lines should be quasi-independent of the atmospheric parameters
because the continuous opacity is predicted to be from photoionization
of carbon from levels only slightly greater in excitation than those
providing the observed lines. These lines have very similar strengths
across the R CrB sample (as expected) but the observed strengths
are markedly weaker than predicted by classical (H-deficient) atmospheres.
After a comprehensive review of a suite of possible explanations for
this `carbon problem', Asplund et al. proposed a flattening of the
temperature gradient; the flattening investigated was assumed to
apply to the entire atmosphere but it might possibly arise as a net
effect of severe stellar granulation. The authors remarked that
`spectra of (H-rich) F supergiants should be studied in attempts to
trace similar effects'. This we have done for a star whose
composition may be rather securely inferred from the RGB (and other)
cluster members.

Although our study might appear to suggest that
the `carbon problem' affects H-rich as well as H-deficient supergiants,
Galactic warm supergiants
both in and
out of the Cepheid instability strip
may be innoculated against the `carbon problem'; the compositions
determined spectroscopically are consistent with the various
lines of evidence independent of spectroscopy that these stars
must have an approximately solar metallicity.
Thus, the parameter space over which the `carbon problem'
affects H-rich supergiants remains to be defined. Overall
metallicity may be a key factor.

\section{Concluding remarks}

Our exploration of the PAGB star in M79 began with the expectation that
either the star might show the canonical effects of the third dredge-up on the
AGB (i.e., notably, a carbon and $s$-process enrichment) or the effects of
dust-gas winnowing or the FIP effect. The exploration was derailed when it
became apparent that an analysis using standard tools of the trade
showed no evidence of expected abundance effects but rather an apparent
0.5 dex deficiency of iron and other elements.  Qualitative considerations
of how standard tools might be modified to effect an upward revision
of the abundances by 0.5 dex led to the idea that the temperature
profile of the PAGB's atmosphere may be flatter than in a classical
atmosphere.

Acquisition and analysis of a superior high-resolution spectrum may
test our proposal concerning the atmospheric structure. It will not
be difficult to improve upon the quality of our spectrum acquired with
a `small' (2.7 m) telescope: high S/N ratio and wavelength coverage are
desirable. Detailed analysis of the spectral energy distribution, particularly
across the Balmer jump may be valuable in assessing the temperature
profile.

Our conclusion that standard model atmospheres are
an inadequate representation of this post-AGB star would not
have been drawn had the star not been a member of a globular
cluster. For a field star, the conclusion would simply have been
that [Fe/H] $=-2$ represented the star. The conclusion also
suggests that abundance analyses of field PAGB stars may need
reconsideration. Construction - empirically or theoretically -
of more realistic model atmosphere will be very challenging but
the task seems important if one is to unravel more completely
the secrets of internal nucleosynthesis, dust-gas winnowing and the
FIP effect that the post-AGB stars hold in the atmospheric
compositions.

\section{Acknowledgments}

This research was begun when Howard Bond and Mike Siegel
unselfishly alerted us to their discovery of the PAGB
star in M79. We thank them both for the opportunity to
analyse this interesting star.
We are especially grateful to Eugenio Carretta for providing
unpublished information on the composition of RGB stars in M79. We
thank Fiorella Castelli for providing bolometric corrections for
stars like the PAGB star in M79 and Bengt Gustafsson and Kjell
Eriksson for offering
comments on the use of spherical rather than plane-parallel atmospheres.
 This research has been supported
in part by the grant F-634 from the Robert A. Welch Foundation of Houston,
Texas.

\end{document}